\def\etal{{et al.\thinspace}}

\def\spose#1{\hbox to 0pt{#1\hss}}
\def\approxlt{\mathrel{\spose{\lower 3pt\hbox{$\sim$}}
        \raise 2.0pt\hbox{$<$}}}
\def\approxgt{\mathrel{\spose{\lower 3pt\hbox{$\sim$}}
        \raise 2.0pt\hbox{$>$}}}

\def\multleft#1{\hbox to size{\vbox {\halign {\lft{##}\cr #1}}\hfill}\par}
\def\multright#1{\hbox to size{\vbox {\halign {\rt{##}\cr #1}}\hfill}\par}

\def\degmark{^\circ}
\def\boxit#1{\vbox{\hrule\hbox{\vrule\kern3pt\vbox{\kern3pt
          #1 \kern3pt}\kern3pt\vrule}\hrule}}

\def\cm{{\rm\thinspace cm}}

\def\erg{{\rm\thinspace erg}}

\def\ph{{\rm\thinspace ph}}
\def\s{{\rm\thinspace s}}

\def\chisq{\hbox{$\chi^2$}}

\def\cntspixps{cts pixel$^{-1}$ s$^{-1}$}

\def\pcmsq{\hbox{$\cm^{-2},$}}

\def\pcmcu{\hbox{$\cm^{-3}\,$}}

\def\ergpcmps{\hbox{$\erg\cm^{-3}\s^{-1}\,$}}
\def\ergpcmsqps{\hbox{$\erg\cm^{-2}\s^{-1}\,$}}

\def\ergps{\hbox{$\erg\s^{-1}\,$}}

\def\pcm{\hbox{$\cm^{-3}\,$}}
\def\pcmsq{\hbox{$\cm^{-2}\,$}}

\def\phpcmsqps{\hbox{$\ph\cm^{-2}\s^{-1}\,$}}

\documentclass{aastex}

\received{}
\accepted{}

\slugcomment{Submitted to the Astrophysical Journal}

\shorttitle{X-rays from NGC 4151}
\shortauthors{Yang, Wilson and Ferruit}

\begin{document}

\title{Chandra X-ray Observations of NGC 4151}

\author{Y. Yang, A. S. Wilson\altaffilmark{1}}

\affil{Astronomy Department, University of Maryland, College Park, MD
  20742; yyang@astro.umd.edu, wilson@astro.umd.edu}

\and

\author{P. Ferruit}

\affil{Observatoire de Lyon, 9 Avenue Charles Andr\'e, Saint-Genis Laval Cedex, F69561, France; 
ferruit@cumulus.univ-lyon1.fr}


\altaffiltext{1}{Adjunct Astronomer, Space Telescope Science
  Institute, 3700 San Martin Drive, Baltimore, MD 21218;
  awilson@stsci.edu}


\begin{abstract}
We present {\it Chandra} X-ray observations of the nearby Seyfert 1.5 
galaxy NGC~4151. The images show the extended soft X-ray emission on the 
several hundreds of pc scale with better sensitivity than previously obtained.
We show that the hard X-ray component ($>2$ keV) is spatially unresolved.
The spectrum of the unresolved nuclear source may be described by a heavily absorbed 
($N_{H} \simeq 3 \times 10^{22}$ \pcmsq), hard power-law ( $\Gamma \simeq 0.3$) plus soft emission from 
either a power-law ($\Gamma \simeq 2.6$) or a thermal ($kT \simeq 0.6$ keV) component. The flux of the high energy component
has decreased from that observed by ASCA in 1993 and the spectrum is much
harder 
($\Gamma \sim 0.3$ in 2000 versus $\Gamma \simeq 1.5$ in 1993).
The large difference between the soft and hard spectral shapes does not favor 
the partial covering or scattering model of the ``soft excess''.
Instead, it is likely that the hard and soft nuclear components represent
intrinsically different X-ray sources.
The stronger nuclear 
emission lines of those seen by the {\it Chandra} HETGS 
spectrum are detected. Spectra of the extended emission to almost 1 kpc NE and SW of the nucleus have 
also been obtained. The spectra of these regions may be described
by either thermal bremsstrahlung ($kT \simeq 0.4-0.7$ keV) or power-law ($\Gamma \simeq
2.5-3.2$) continua plus 3 emission lines.  There is 
an excellent correlation between the extended X-ray and [O {\sc iii}]$\lambda 5007$ line emissions. 
We discuss the nature of the extended X-ray 
emission. Because there is no extended electron-scattered hard X-ray emission, 
an upper limit to the electron scattering column can be obtained. This upper
limit is much too low for the soft X-rays to be electron scattered nuclear radiation, 
unless the nucleus radiates soft X-rays much more strongly towards 
the extended regions than towards Earth, a situation we consider unlikely. We
favor a picture in which the extended X-ray emitting gas is heated {\it in situ}
by the nuclear radiation. Some of the X-rays may originate from a hot phase 
which
confines the warm ionized gas seen optically, although X-ray emission
produced via photoionization by the nucleus is also likely. A faint, probably
background, compact X-ray source lies $\simeq$ 2\farcm2 from the nucleus
to the SW, approximately along an extension of the extended SW X-ray and
[OIII] emission.

\end{abstract}




%

\newpage
\section{Introduction}

Because of its proximity (13.2 Mpc for $H_{0}=75$ km s$^{-1}$ Mpc$^{-1}$, 
$1\arcsec = 64$ pc), the Seyfert 1.5 galaxy 
(Osterbrock \& Koski 1976) NGC~4151 is one of the best studied among the class. 
X-ray observations show that NGC~4151 is a modest luminosity ($L_{2-10 keV}\simeq 2-20 \times 10^{42}$ \ergps) active
galactic nucleus (AGN) with a highly variable and hard continuum.
There is evidence for a spectral-flux correlation above 2 keV (Perola \etal~1986; Yaqoob \& Warwick 1991; 
Yaqoob \etal~1993). The soft X-ray continuum of NGC~4151 is characterized by a prominent excess of X-ray 
emission above an extrapolation of the absorbed, higher energy power-law component to lower energies. This ``soft excess'' has 
been  described in terms of partial covering, a warm absorber, scattered nuclear radiation and additional spectral components 
(e.g. Holt \etal~1980; Yaqoob \& Warwick 1991; Weaver \etal~1994a, b). 

{\it Einstein} and {\it ROSAT} observations identified extended soft X-ray emission
that appears to be associated with ionized gas visible optically (Elvis, Briel \& Henry 1983; Morse \etal 1995). 
This extended emission must account for at least part of the ``soft excess''. Ogle \etal~(2000) have recently obtained  
a 48 ks observation of NGC~4151 with the High-Energy Transmission Grating Spectrometer (HETGS) on board 
the {\it Chandra X-ray Observatory}. Their observation provided the first high resolution X-ray spectroscopy
of NGC~4151 and also a zero-order image. Strong, narrow X-ray emission lines are seen in the spectrum. Evidence
of both photoionized and collisionally ionized gas was found. However, their zero-order image has only modest 
signal-to-noise due to the limited transmission of the gratings. 

We thought it worthwhile to obtain an exposure without the gratings 
in place in order to obtain a sensitive image of the extended emission. 
Given the dependence of the HETGS zero-order
transmission on energy, our $\sim 30$ ks direct image has higher sensitivity than the 
zero-order image by Ogle \etal (2000) for all energies below $\simeq 6$ keV. 
  
In this paper, we present a {\it Chandra} Advanced CCD Imaging Spectrometer (ACIS) observation
of NGC~4151. Observations are described in Section 2. Data analysis and results are presented in 
Section 3, where we present the best X-ray image of NGC~4151 to date and low resolution spectra of both 
the nucleus and extended regions. The X-ray image is also compared with optical and radio images. 
Implications of the observation are discussed in Section 4. The detection of an off nuclear source 
about 2\farcm2 SW of the nucleus of NGC~4151 is described in the Appendix.

\section{Observations and Effects of Pile Up}
The observations of NGC~4151 were obtained in two phases. First, to evaluate 
the effects of pile up\footnote{For bright sources, two or more photons may fall in 
the same pixel during one frame time interval and the 
photons are then detected as a single
event. This ``pile up'' effect results in an underestimated count rate
and an incorrect spectrum.} in the bright inner regions of NGC~4151, we took a preliminary 4.3~ks 
observation on Dec. 4 1999 in so called ``alternating  mode'', in which single exposures 
with a 0.1~s frame time were alternated 
with two exposures with a 0.4~s frame time.  The observation showed that 
the nucleus is significantly piled up in the 0.4~s, but not 
in the 0.1~s, frame time observations. We also inferred that the bright inner regions of the galaxy 
would be strongly affected by pile up in the planned
long exposure, which would use the standard 3.2~s frame time. We thus decided to obtain another, longer
observation in ``alternating mode'', in addition to the long observation, to mitigate the effect of pile up
in the inner regions.   Therefore, the second phase of observations
consisted of a $\simeq$ 26~ks observation with 3.2~s frame time and a $\simeq$ 7~ks
observation in ``alternating mode'', including both 0.1~s and 0.4~s exposures.  
Observations are summarized in Table~1. In the $\simeq 26$ ks observation, CCDs I2, I3, S1, S2, S3
and S4 were read out, though all of the detected X-ray emission from
NGC 4151 is on S3. In both ``alternating mode '' exposures, only chip S3 was read out. 

The fraction of bad grades (i.e. the ratio of the number of counts with ASCA grades 1+5+7 [defined in the {\it the Chandra 
Proposers' Observatory Guide}, section 6.3] to the total number of counts)
was found to be a good indicator of the degree of pile up in any given pixel through comparison with the event rates in the different frame time observations.
Regions with a large fraction (nominal value $>10\%$) of bad grades were 
considered to be piled-up. As already found in the preliminary observation,
the 0.1~s frame time data were not piled-up, with less than 6\% of the grades
in the brightest pixel being bad. The 0.4~s and 3.2~s frame time data were 
found to suffer from pile-up out to 1\farcs3 and $3\arcsec$ from the nucleus, respectively.

All of the analysis was initially done with the ``old'' response functions
(``fefs''), and  data 
were reduced and analysed with  {\sc ciao} v1.1.5 and {\sc xspec} v11.0.1.
The observations were screened for high
background count rates
and aspect errors by following the {\it Chandra Science Threads}.
After we received the referee's report, new versions
of response function (CALDB 2.7) were
available. We reanalysed some of the data with
these new response functions
and with the latest version of {\sc ciao} -- v2.1.2.  
In general, the change in derived
parameters (such as absorbing column, photon index, bremsstrahlung temperature and emission line 
properties) were small. We did not reanalyse the 0.1~s frame time nuclear 
observations with the 
new response functions because our results agree well with the concurrent HETGS observations of the nucleus by 
Ogle et al. (2000), which have much higher spectral resolution and are thus superior for detection of lines. 
We have, however, added 0.2 keV to the line energies below 1 keV from the 0.1~s frame time observation since
there is a systematic error of this magnitude when using the old response 
functions. The results presented for the 0.4s frame time nuclear observations  
and the 3.2s frame time observations of the extended emission employed the
new response functions.
  
\section{Data Analysis and Results}

\subsection{Morphology of X-ray emission}

The soft X-ray (0.3--2.5 keV) images derived from the 0.1~s, 0.4~s and 3.2~s 
frame time exposures are shown in Fig. 1. The ``hole'' at the position
of the nucleus in the 3.2~s frame time image (Fig. 1c) is due to pile-up.
The X-ray emission of NGC 4151 comprises a bright unresolved nucleus and resolved 
extended regions. The south-west part of the extended emission extends 
as far as $14\arcsec$ ( $\simeq 900$ pc) from the nucleus along PA $ \simeq 233\degmark$. The north-east part 
of the extended emission extends to $\approx 11\arcsec$ ($\simeq 700$ pc) from the nucleus along PA $\simeq 75\degmark$.  
The structure of the extended emission appears to be knotty rather than smooth.  A moderately bright, compact knot is 
clearly visible $6\arcsec$ ($\simeq 380$ pc) south-west of the nucleus (Fig. 1c). A contour map 
representation of the extended emission is shown in Fig. 2. Our images are in good agreement with the 
HETGS zero-order image obtained by Ogle et al. (2000), but are more sensitive.    

The 2--9 keV emission is not resolved in our observations, as can be seen by
comparing the radial profile of the observations in this band with the point spread function (PSF). 
In Fig. 3a, the 5.5 keV model PSF without 
aspect error is plotted with the observed profiles of the 2--9 keV emission. 
The 0.4--2 keV emission is, on the other hand, resolved along the northeast-southwest direction, 
but is not (or is marginally) resolved along the northwest-southeast direction (Figs. 3b, c, d).     

\subsection{X-ray Spectra of the Nuclear Region}

We extracted spectra of the nucleus from a circular region with radius $3\arcsec$ in the 0.1~s and 0.4~s 
frame time observations. The total number of counts in the spectra are 2033 
and 10690 in the
0.1~s and 0.4~s frame times, respectively. 
For both observations, the background was taken from a region between two 
ellipses centered on the nucleus. The large ellipse had a semi-major axis of 
17\farcs 5 along PA $27\degmark$ (the column direction of CCD read out) and 
the smaller ellipse had a semi-major axis of 12\farcs5 in the direction of  PA $= 52\degmark$. 
This rather oddly-shaped background region allowed all the background 
events to be extracted from the same CCD node. 

The extracted spectra were grouped to a minimum of 20 and 30 counts per bin in the 0.1~s and 0.4~s frame time data,  
respectively, to allow use of $\chisq $ statistics. The spectral resolution of ACIS-S is $\sim 0.1$ keV at 1 keV.
For the bright nuclear region, the spectral resolution was not significantly degraded by this grouping. 
Because of the poor calibration at low energies and high background at high energies, channels 
below 0.3 keV and above 9 keV were ignored when modeling the spectra.

The emission from the brightest part of NGC 4151 originates in the 
nucleus and the unresolved emission line regions. 
At least two components are needed to describe the continuum of the spectra from this region: 
a hard power law absorbed by an equivalent hydrogen column density of $N_{H}=(2-3)\times10^{22}$ \pcmsq  
and a soft component that can be represented by either an absorbed power law or a thermal bremsstrahlung  model. 
The 0.1~s frame time spectra were fitted with these two component continuum models (Table 2) 
plus emission lines (Table 3). The model 2--9 keV flux is $4.8 \times 10^{-11}$ \ergpcmsqps and the photon index of the
hard component  $\Gamma \simeq 0.32 $.
The soft X-ray spectrum below 2 keV can be described as a power law with photon index $\Gamma \sim 2.6$ (Table 2, Fig. 4a).  
An alternative description in terms of thermal bremsstrahlung gives a temperature $\sim 0.57$ keV (Fig 4b). 
However, the emission cannot be described by solar abundance, collisionally ionized hot gas. 
The best fit double power law parameters agree with those obtained in a concurrent HETGS 
observation (Ogle \etal~2000). Previous {\sc ASCA} observations (Weaver \etal~1994b) gave a
2--10 keV flux of  $11-22 \times 10^{-11}$ \ergpcmsqps, a photon index of  
$\Gamma \sim 1.5$ and an absorbing column of $2-5 \times 10^{22}$ \pcmsq.
Weaver et al.'s absorbing column is about the same as found in  our observation and that of 
Ogle \etal (2000), while the photon index and flux from ASCA
are significantly larger, indicating variability of the hard component.
Our observations suggest that the hard component of NGC~4151 is at its 
lowest state since 1984 ({\it EXOSAT} observations yielded a low 2--10 keV flux
of $3.6 \times 10^{-11}$ \ergpcmsqps,
[Pounds et al. 1986]). The very hard power law index found in the present
observations with {\it Chandra} 
is unusual for a Seyfert 1 galaxy (Mushotzky 1984). 

Lines were identified by adding narrow Gaussian features to the continuum model with the goal of reducing 
the $\chi ^{2}$ value of the fit.  
For all cases except the combined O {\sc vii} triplet, the intrinsic line width was set to zero. 
Both 0.1~s and 0.4~s frame time data were used. Though piled-up, the continuum
of the 0.4~s frame time data  between 0.4 to 7 keV can still be described by a double power law model reasonably well
(reduced $\chi^{2}=180.8$ for 179 degrees of freedom), with a rather hard photon index 
($\Gamma = -0.14^{+0.08}_{-0.07}$) above 2 keV 
(Fig. 4c). This value of $\Gamma$ is, however, too low as a result of the pile up. The low spectral resolution of
the ACIS instrument does not allow us to resolve all the lines seen in the grating observation
(Ogle \etal 2000), but the stronger lines are detected. The best fits to the spectra are shown in Fig. 4, while the lines 
identified are listed in Tables 3 and 4. The errors represent the 90\% confidence range for a 
single interesting parameter. One feature in the 0.4~s frame time spectrum, which probably results from the rapid 
change of effective area with energy near 2 keV, is also listed in 
Table 4 and labeled as ``artefact''. The line fluxes from the 0.1~s frame time image agree well with those 
found in the HETGS observation (Table 3). The line fluxes found from the 0.4~s frame time data tend to be lower than those 
obtained from the grating observation (see Table 4), probably as a result of pile up.

\subsection{X-ray Spectra of the Extended Emission}

For the extended emission, we used the 3.2 s frame-time observation and extracted spectra from  
non-piled-up regions more than $3\arcsec$ from
the nucleus to both NE and SW.   
The spectrum of the NE region was extracted from a 12\farcs7 $\times$ 8\farcs6 
rectangle centered 7\farcs3 northeast of the nucleus 
($\alpha = 12^{h} \thinspace 10^{m} \thinspace 33^{s}.0$, 
$\delta = 39\degmark \thinspace 24^{\prime} \thinspace 26\farcs7$) with
the longer side in position angle $144 \degmark$. 
The spectrum of the SW region 
was extracted from a 12\farcs3$\times$7\farcs7 rectangle centered 9\farcs2 southwest of the nucleus 
($\alpha = 12^{h} \thinspace 10^{m} \thinspace 31^{s}.8$, 
$\delta = 39\degmark \thinspace 24^{\prime} \thinspace 17\farcs9$ ) with the longer side in 
position angle $234\degmark$ (Fig. 5). The total number of counts in the NE and SW regions are 1120 and 1776, 
respectively. The background region was the same as in section 3.2.    
The data were grouped to a minimum of 20 counts per bin in each region. The spectral resolution 
above 1 keV was significantly reduced by this grouping, especially for the NE region. As before, channels below 0.3 
and above 9 keV were ignored in spectral modelling.

The extended soft X-ray continuum emission can be modeled by a single bremsstrahlung or power 
law model with photo-electric absorption (Table 5).  
The emission observed above 2 keV is the unresolved nuclear emission which has been spread out by the
mirror PSF (see section 3.1 and Fig. 3a), and are hence modeled with an absorbed power law with photon index
and absorbing column density identical to the nuclear value (section 3.2). 
The absorbing column densities for the extended regions are not well
constrained with the present data and 
were ``frozen'' to the Galactic value  (N$_{\rm H,Gal} = 2.19 \times 10^{20}$ \pcmcu, Murphy \etal~1996).

It would be interesting to examine the spectral index as a function of distance 
from the nucleus. However, the low count rate in the extended region
makes it hard to obtain such spectra. To obtain a more qualitative picture,
the ``hardness ratio'' of the soft X-ray band (defined as the ratio of background
subtracted count rates in the 0.8--2.0 keV and 0.4--0.8 keV bands) as a function of
distance from the nucleus has been obtained for the SW region from the 3.2 s frame time 
observation (Fig. 6). No significant change of hardness ratio is seen between $4\arcsec$ and $14\arcsec$ from the
nucleus.     

Spectral lines were identified with the same technique as for the nucleus, and the results are shown
in Table 6. He-like O and Ne triplets are seen in both the SW and NE regions. 
The 0.72--0.74 keV feature observed in both regions could be a blend of lines 
of H-like O and radiative recombination continuum 
features of O {\sc vii}. The observed spectra are compared with the 
continuum plus emission-line models in Fig. 7.

\subsection{Comparison with observations at other wavelengths}

An [O {\sc iii}]$\lambda 5007$ contour map (P\'{e}rez-Fournon \& Wilson 1990) is overlaid on the  3.2~s frame time 
0.3--9 keV image in Fig 8. Extended X-ray emission is well correlated with the forbidden line emission 
to both NE and SW of the nucleus. The X-ray knot $6\arcsec$ ($\simeq 380$ pc) to the SW of nucleus is also seen in 
the [O {\sc iii}] image, but with a $\sim 1\arcsec$ displacement. 

In Fig. 9, a radio contour map from MERLIN observations (Mundell \etal~1995) is overlaid on our 0.1~s frame time
{\it Chandra} image. The radio jet does not align with the soft X-ray extension. The angle between the two directions 
is $\sim 12\degmark$.

\section{Discussion}           
\subsection{Partial covering and Scattering}
One of the early suggestions for the origin of  ``soft excess'' is the partial covering model (Holt \etal~1980),
in which the continuum from the nucleus is covered by a ``clumpy'' absorber.  
Previous studies show that a covering factor of 0.7 to 0.97 can account for all
or a significant part of the ``soft
excess'' (e.g. Weaver et al. 1994b, Warwick et al. 1995).
However, in the two concurrent 
observations presented by Ogle \etal ~(2000) and this paper, the photon indices for the hard and  soft component 
were found to be $\simeq 0.3$ and $\simeq 2.6$, respectively.  The 0.1--2 keV
flux from our observation is $\sim 4 \times 10^{-12}$ \ergpcmsqps, which is 
close to the lowest value seen in the
{\it ASCA} observations of $\sim 5 \times 10^{-12}$ \ergpcmsqps(Weaver et al.
1994b). This shows no significant 
variability in the soft component, in contrast to the large variability in the
hard component. These results makes a partial covering model 
unlikely. 
The same argument applies to the notion that the soft emission is  
Thomson scattered nuclear radiation, as already noted by Ogle \etal (2000). 
Instead, it is most likely that the soft and hard components 
are intrinsically different X-ray sources.

The Thomson scattering model for the extended soft 
emission can be evaluated using the soft- and hard-band profiles (Fig. 3). The hard-band profile is consistent with the mirror PSF, indicating there is no detectable intrinsically extended hard emission. This allows an upper limit to the fractional brightness of any hard nuclear emission which is Thomson scattered by the extended gas to be obtained. Because Thomson scattering is wavelength independent, this upper limit can be compared with the brightness of the extended, soft emission, expressed as a fraction of the soft nuclear emission. Using the 0.1~s frame time observations in Fig. 3a, the 2--9 keV brightness within $1\arcsec$ of the nucleus is 0.12 \cntspixps~
and the intrinsic 2--9 keV brightness between $4\arcsec$ and $5\arcsec$ from the nucleus is $< 3.9 \times 10^{-5}$ \cntspixps, which represent an upper limit to the Thomson scattered hard radiation. The corresponding numbers for the 0.4--2 keV emission are 0.13 \cntspixps~for the nucleus and $2.6 \times 10^{-4}$ \cntspixps~between $4\arcsec$ and $5\arcsec$ from the nucleus in the SW region (Fig. 3c). Defining $R \equiv$(brightness between $4\arcsec$ to $5\arcsec$ from the nucleus)/(brightness within $1\arcsec$ of the nucleus), we have $R_{2-9~keV} < 3.2 \times 10^{-4}$ and $R_{0.4-2~keV} = 2 \times 10^{-3}$. The argument indicates that the extended soft emission is not Thomson scattered nuclear radiation as long as the intensity of the nuclear radiation radiated towards the putative scattering region is equal to that radiated towards Earth. The nuclear 2--9 keV radiation radiated towards the extended region may be stronger than that radiated in  our direction because the latter is absorbed. Such an effect decreases $R_{2-9~keV}$ and increases the discrepancy between $R_{2-9~keV}$ and $ R_{0.4-2~keV}$. If the soft nuclear emission is absorbed by more than the low column we obtained by fitting the spectrum, it is possible that more soft radiation is radiated from the nucleus towards the extended emission than towards us. This effect would reduce $R_{0.4-2~keV}$ and could potentially remove  the discrepancy between $R_{0.4-2~keV}$ and
$R_{2-9~keV}$. While such cannot be ruled out, we consider it unlikely that the extended X-ray emission is Thomson-scattered nuclear radiation.     

\subsection{A Two Phase Model for the Extended X-ray Emission}

The close association between the extended soft X-ray emission and the optical line emission 
in NGC 4151 suggests that the X-rays may arise from hot gas in pressure 
equilibrium with the extended narrow line region (ENLR) (Heckman \& Balick 1983,
Morse \etal~1995). Gas of $10^{6}-10^{8}$ K is needed to confine the cooler 
gas clouds (Krolik, McKee \& Tarter 1981, hereafter KMT;  Krolik \& Vrtilek 1984). 
Grating observations with the {\it Chandra} HETGS show clear signatures of both 
photoionization and  collisional ionization in the X-ray emitting gas in the narrow line region. 
This indicates at least two phases exist in the plasma, one being hot
($T \simeq 10^{7}$ K) and collisionally ionized and the other warm ($T \geq 10^{4}$ K) 
and photoionized (Ogle \etal 2000).
 
Penston \etal~(1990) derived emissivity-weighted estimates for the electron
density and electron temperature in the SW optical ENLR of $n_{e} \simeq 220 $ \pcmcu 
and $T_{e} \simeq 14130$ K, corresponding to a pressure of $n_{e}T_{e}=3.1 \times 10^{6}$ \pcmcu K.
If the warm and hot clouds are in a rough pressure equilibrium, the density of the hot phase gas 
at a temperature of $T_{h}=4.8 \times 10^{6}$ K (the bremsstrahlung temperature in the SW regions) should be
$n_{h} \simeq 0.6$ \pcmcu. Assuming the hot gas in the observed SW region uniformly fills a cylinder with a radius of 
246 pc ($\simeq$ 3\farcs85) and a height of 787 pc ($\simeq$ 12\farcs3 ), the number density inferred 
from the bremsstrahlung description of the continuum of the SW radiation (see Table 5) is found to be 
$n_{h} \simeq 0.5$ \pcmcu. This is close to that needed for pressure balance.
Using the two power-law
spectrum obtained from the 0.1~s frame time observation (section 3.2) and adopting a spectral turn-over near 100 keV
(Maisack \etal~1993), we obtain a photoionizing luminosity $L_{13.6~eV - 100~keV}= 6.2 \times 10^{43}$ \ergpcmps.
 The ionization parameter (defined as $\Xi \equiv L/4 \pi ckTnr^{2}$, KMT) 
for the hot gas measured at the center of the SW region ($ r= 589$ pc) is then 0.19. This combination of temperature 
and ionization 
parameter lies in the general area of the equilibrium curves of $T$ 
versus $\Xi$ calculated by KMT, although the equilibrium would be unstable.
We note that the temperature
derived from the bremsstrahlung model is much lower than the Compton temperature ( $T \geq 10^8$  K, see KMT).
Given our low spectral resolution, we cannot tell whether the emission lines we see from the 
extended region are collisionally ionized
or photoionized. Ogle \etal~(2000) identified photoionized 
N {\sc vii}, O {\sc vii}, O {\sc viii}, Ne {\sc ix} and Ne {\sc x} in the 
nuclear region from the narrowness of their radiative recombination continua. 
We see lines (some tentatively) of O {\sc vii}, O{\sc viii},
Ne {\sc ix} and Ne {\sc x} from the extended regions (Table 6). If these species
are created by photoionization, the ionizing
parameter lies in the range $1.6<log~\xi<2.2$, where $\xi \equiv L/nr^{2}$ (Kallman \& McCray 1982). 
Using the ionization luminosity
given above, and a typical radius of 500 pc, these values of $\xi$ imply $0.2 < n< 0.7$ \pcm. 
The temperature of this gas would be 
$\simeq 10^{4.5}$ K according to Kallman \& McCray (1982). Higher spectral resolution 
observations are needed to determine whether these
lines are photoionized in such a warm gas or collisionally ionized in a hotter one.

\section{Conclusion} 
We have obtained high sensitivity X-ray imaging-spectroscopy of NGC~4151 with the {\it Chandra X-ray Observatory} .  
The soft X-ray emission (below 2 keV) is spatially resolved and extends as far 
as $\sim 900$ pc to the SW and $\sim 700$ pc to the NE of the nucleus. There is a close 
correlation between the extended soft X-ray and [O {\sc iii}]$\lambda 5007$ emissions  
on the hundreds of pc scale. The 0.4--9 keV  spectra of the nucleus and the extended 
regions to the SW and NE of the nucleus have been obtained. The X-ray emission 
above 2 keV is spatially unresolved and is well described by an absorbed power law. 
The spectrum above 2 keV is harder and the flux lower than in observations made with ASCA 
in 1993 (Weaver \etal 1994b). The spectrum below 2 keV, on the other hand, is softer 
than found previously. The large difference between the soft and hard spectral shapes does not 
favor the partial covering or scattering model of the ``soft excess''.
Instead, it is likely that the hard and soft nuclear components represent 
intrinsically different X-ray sources.
We have shown that a model in 
which the extended emission is electron-scattered nuclear radiation is extremely implausible. 
With the temperature ($\sim 4.8\times 10^{6}$ K) inferred from a bremsstrahlung interpretation 
of the X-ray spectrum, this hot gas is close to pressure equilibrium with the gas which radiates the 
optical line emission. While the extended emission is certainly thermal, our observations do not permit a clear
distinction between a photoionized and collisionally ionized gas. It is likely that both components are 
present, given the findings of Ogle \etal~(2000) for the nuclear region.

We thank Andy J. Young, Patrick L. Shopbell and David A. Smith for help with using CIAO,
XSPEC and WIP in analyzing the data. We are also grateful to Carole G. Mundell for providing
the MERLIN map in the numerical form. This research was partially supported by NASA grant
NAG 81027, by HST grant GO 7275,  and by the Graduate School of the University of Maryland.

\section*{APPENDIX: An Off Nuclear X-ray Source}
An unresolved X-ray source
($\alpha(2000.0) =  12^{h} \thinspace 10^{m} \thinspace 22^{s}.37$, 
$\delta(2000.0) = 39\degmark \thinspace 23^{\prime} \thinspace 17\farcs0$) 
$\simeq$2\farcm2  SW of the nucleus (P.A. $\simeq 241\degmark $) 
of NGC~4151 has been detected in the 3.2~s frame time observation. 
The total counts within a $3\arcsec$ radius centered on the point source
is 249. We obtained a spectrum of the 
source (Fig. 10). The spectrum can be well described by a power law absorbed by the Galactic
column density (N$_{\rm H,Gal} \simeq 2.19 \times 10^{20} $ \pcmcu, Murphy \etal~1996), with 
$\chisq = 27.2$ for 25 degrees of freedom. The best fit photon index 
$\Gamma = 1.7^{+0.1}_{-0.2}$ and the model flux for the 
0.3--5 keV energy range is 
$ 6.8 \times 10^{-14}$ \ergpcmps, corresponding to a luminosity of 
$1.4\times10^{39}$ \ergps, 
if the source is associated with 
NGC~4151. This source lies approximately along the extension of the SW extended
emission to larger radii. There is a weak optical object nearby at the limit of the 
DSS with an estimated POSS-E magnitude of 
$\approx 20$. It is unclear whether this X-ray source is physically 
associated with NGC 4151. 
Similar coincidences have been seen in Mkn 3, Pictor A and NGC 4258, 
each of which has a faint, compact X-ray source close to the extension of the 
radio jets (Morse \etal~1995; Wilson, Young \& Shopbell 2001;
Wilson, Yang \& Cecil 2001).


\vfil\eject\clearpage
\begin{deluxetable}{ccccc}
\tabletypesize{\footnotesize}
\tablewidth{0pt}
\tablecaption{Observation Summary}
\tablecolumns{5} \tablehead{\colhead{Obs ID} & \colhead{Sequence \#} & 
\colhead{Observation date} & \colhead{Frame time} &
\colhead{Exposure\tablenotemark{a}}} \startdata
347 & 700019 & 12/04/99 11:29:54 & 0.1 sec & 296.8 sec\\
347 & 700019 & 12/04/99 11:29:54 & 0.4 sec & 2374.3 sec  \\
348 & 700020 & 03/15/00 10:51:40 & 3.2 sec & 26.2 ksec \\
372 & 700198 & 03/06/00 13:27:39 & 0.1 sec & 771.1 sec \\
372 & 700198 & 03/06/00 13:27:39 & 0.4 sec & 6168.8 sec \\
\enddata
\tablenotetext{a}{Total good time with dead time correction. }
\end{deluxetable}

\vfil\eject\clearpage
\begin{deluxetable}{cccccccc}
\tabletypesize{\footnotesize}
\tablewidth{0pt}
\rotate 
\tablecaption{Spectral Models of the Nucleus
  From the 0.1~s Frame Time Observations \label{tab:nuc_cont}}
\tablecolumns{8} \tablehead{\colhead{Model\tablenotemark{a}} &
  \colhead{$N_{H}$ (Soft Comp.)} & \colhead{kT\tablenotemark{b} / $\Gamma_{1}$} & 
  \colhead{$K_{1}$\tablenotemark{c}} & \colhead{$N_{H}$ (Hard Comp.)} & 
  \colhead{ $\Gamma_{2}$}&
  \colhead{$K_{2}$\tablenotemark{c} } &
  \colhead{\chisq~/ d.o.f.}  \\ 
  \colhead{} & \colhead{[$\times 10^{20}\pcmsq$]} & 
  \colhead{ } & \colhead{} & \colhead{[$\times 10^{22}\pcmsq$]} & \colhead{ } 
   & \colhead{ } & \colhead{ }} \startdata

PL+PL+LINES & $2.9^{+0.7}_{-0.7}$ & $2.6^{+0.1}_{-0.1}$ &
$\left( 7.8^{+0.4}_{-0.5} \right) \times 10^{-4}$ & $3.1^{+0.4}_{-0.3}$ & $0.32^{+0.05}_{-0.12}$ &
$\left( 1.4^{+0.1}_{-0.0} \right) \times 10^{-3}$ & 66.3/65\\

BR+PL+LINES & $2.0^{+0.7}_{-0.7}$ & $0.57^{+0.7}_{-0.8}$ &
$\left( 2.9^{+0.9}_{-0.5} \right) \times 10^{-3}$ & $2.6^{+0.5}_{-0.5}$ & $0.32^{+0.10}_{-0.21}$ &
$\left( 1.4^{+0.3}_{-0.5} \right) \times 10^{-3}$ & 70.4/69\\

\enddata

\tablenotetext{a}{Model abbreviations are:  PL = power law, BR = bremsstrahlung, LINES = emission lines (see Table 3 for
	line parameters).}
\tablenotetext{b}{Unit: keV}
\tablenotetext{c}{Model normalization.
  For the bremsstrahlung model $K_{\rm Brem} = 3.02 \times 10^{-15}
  \int n_e n_I dV / (4 \pi D^2)$, where $n_e$ is the electron density, $n_I$ is the ion density and 
  $D$ is the distance to the source (all in cgs units). For the power law model $K_{\rm PL}
  = \phpcmsqps {\rm keV}^{-1}$ at 1 keV.}

\end{deluxetable}

\vfil\eject\clearpage
\begin{deluxetable}{ccccccc}
\tabletypesize{\footnotesize}
\tablewidth{0pt}
\rotate
\tablecaption{Emission Lines From the Nucleus: 0.1~s Frame Time Observations
\label{tab:nuc_lines}}
\tablecolumns{7} 
\tablehead{\colhead{Model\tablenotemark{a}} & \colhead{Energy} & \colhead{Line} & \colhead{Observed Energy} 
& \colhead{$K$\tablenotemark{b}}  & \colhead{$EW$\tablenotemark{c}} &\colhead{ Flux\tablenotemark{d}} \\ 
 \colhead{ }& \colhead{[keV]} & \colhead{} & \colhead{[keV]} & \colhead{}  & \colhead{[eV]} & \colhead{}} \startdata

 I & 6.40 & Fe I K${\alpha}$ & $6.37^{+0.04}_{-0.06}$ & 
  $\left(1.4^{+0.8}_{-0.9} \right) \times 10^{-4}$ &  $182$ & $1.8 \times 10^{-4}$  \\

 I & 0.87, 0.90, 0.91 \& 0.92 & O {\sc viii} RRC, Ne {\sc ix} triplet & $0.91^{+0.01}_{-0.02}$ & 
  $\left(8.7^{+3.5}_{-4.1} \right) \times 10^{-5}$ &  78.0 & $7.4 \times 10^{-5}$ \\

 I & 0.74 \& 0.77 \& 0.82 & O {\sc vii} RRC \& O {\sc viii} Ly${\beta}$ Ly$\gamma$ & $0.76^{+0.06}_{-0.05}$ & 
  $\left(4.0^{+6.7}_{-4.0} \right) \times 10^{-5}$ &  21.7 & $ 8.1 \times 10^{-5}$ \\

 I\tablenotemark{e}  & 0.561, 0.568 \& 0.574 & O {\sc vii} triplet & $0.58^{+0.03}_{-0.04}$ & 
  $\left(2.6^{+0.5}_{-0.3} \right) \times 10^{-4}$ & 71.0 & $4.3 \times 10^{-4}$   \\

\hline

 II & 6.40 & Fe I K${\alpha}$ & $6.36^{+0.04}_{-0.03}$ & 
  $\left(1.3^{+0.7}_{-0.6} \right) \times 10^{-4}$  & $181$ & $1.8 \times 10^{-4}$ \\

 II &0.87, 0.90, 0.91 \& 0.92 & O {\sc viii} RRC, Ne {\sc ix} triplet & $0.91^{+0.01}_{-0.01}$ & 
  $\left(1.1^{+0.2}_{-0.4} \right) \times 10^{-4}$  & 105. & $7.4 \times 10^{-5}$ \\

 II & 0.74 \& 0.77 \& 0.82 & O {\sc vii} RRC \& O {\sc viii} Ly${\beta}$ Ly$\gamma$ & $0.76^{+0.01}_{-0.03}$ & 
  $\left(8.7^{+3.2}_{-3.8} \right) \times 10^{-5}$  & 50.7 & $8.1 \times 10^{-5} $  \\
 
 II & 0.65 & O {\sc viii} Ly$\alpha$ & $0.62^{+0.02}_{-0.01}$ & 
 $\left(1.6^{+0.6}_{-0.5} \right) \times 10^{-4}$ & 55.2 & $1.0 \times 10^{-4}$  \\

 II & 0.561, 0.568 \& 0.574 & O {\sc vii} triplet & $0.53^{+0.07}_{-0.06}$ & 
  $\left(1.7^{+0.6}_{-0.7} \right) \times 10^{-4}$ & 37.4 & $4.3 \times 10^{-4}$  \\

\enddata
\tablenotetext{a}{Model I: bremsstrahlung + power law + emission lines; 
	Model II: two power laws + emission lines.}

\tablenotetext{b}{$K = {\rm total } \phpcmsqps$ in the line.}

\tablenotetext{c}{Equivalent Width}

\tablenotetext{d}{Total $\phpcmsqps$ in the line from HETGS observations by Ogle \etal~(2000)}

\tablenotetext{e}{Modelled as a single line with width $0.05 \pm 0.03$ keV; this width is probably not real in view of present calibration uncertainties.}

\end{deluxetable}

\vfil\eject\clearpage
\begin{deluxetable}{cccccc}
\tabletypesize{\footnotesize}
\tablewidth{0pt}
\rotate
\tablecaption{Emission Lines From the Nucleus: 0.4~s Frame Time Observations
\label{tab:nuc_lines}}
\tablecolumns{6} \tablehead{\colhead{Energy} & \colhead{Line} & \colhead{Observed energy} 
& \colhead{$K$\tablenotemark{a}}  & \colhead{$EW$\tablenotemark{b} } &\colhead{Flux\tablenotemark{c} }\\ 
 \colhead{[keV]} & \colhead{} & \colhead{[keV]} & \colhead{}  & \colhead{[eV]} & \colhead{}} \startdata

 6.40 & Fe I K${\alpha}$ & $6.43^{+0.09}_{-0.11}$ & 
  $\left(5.9^{+2.0}_{-3.0} \right) \times 10^{-5}$ &  $92.6 $ & $1.8 \times 10^{-4}$  \\

 2.25 & Artefact & - & - & - & - \\

 1.73, 1.74, 1.84 \& 1.88 &Si {\sc I} K${\alpha}$, Mg {\sc xii} Ly${\beta}$, Si {\sc xiii} f, r & $1.81^{+0.03}_{-0.02}$ & 
  $ \left( 2.1^{+0.6}_{-0.8} \right) \times 10^{-5} $ & 146. & $4.0 \times 10^{-5}$ \\

1.33, 1.35 \& 1.36 &  Mg {\sc xi} f, r, Ne {\sc x} RRC & $1.38^{+0.03}_{-0.03}$ & 
  $ \left( 7.1^{+4.1}_{-4.0} \right) \times 10^{-6} $ & 25.7 & $2.9 \times 10^{-5}$ \\

 1.02  & Ne {\sc x} Ly${\alpha}$ & $1.07^{+0.02}_{-0.03}$ & 
  $ \left( 2.0^{+1.0}_{-0.9} \right) \times 10^{-5} $ & 39.9 & $2.0 \times 10^{-5}$ \\

 0.90, 0.91 \& 0.92 & Ne {\sc ix} triplet & $0.92^{+0.01}_{-0.01}$ & 
  $\left(6.9^{+1.5}_{-1.5} \right) \times 10^{-5}$ &  100. & $6.2 \times 10^{-5}$ \\

 0.74, 0.77 \& 0.82  & O {\sc vii} RRC, O {\sc viii} Ly${\beta}$, Ly${\gamma}$ &
  $0.81^{+0.12}_{-0.13}$ &            
  $\left(5.0^{+1.6}_{-1.5} \right) \times 10^{-5}$ &  54.5 &  $8.1 \times 10^{-5}$\\

 0.65 \& 0.66 & O {\sc viii} Ly${\alpha}$, N {\sc vii} RRC    & $0.67^{+0.05}_{-0.02}$ & 
  $ \left( 8.9^{+2.4}_{-2.7} \right) \times 10^{-5} $ & 60.8 & $1.2 \times 10^{-4}$ \\

 0.561, 0.568 \& 0.574 & O {\sc vii} triplet & $0.56^{+0.02}_{-0.01}$ & 
  $\left(3.5^{+1.0}_{-0.5} \right) \times 10^{-4}$ & 166 & $4.3 \times 10^{-4}$  \\

\enddata

\tablenotetext{a}{$K = {\rm total} \phpcmsqps$ in the line.}

\tablenotetext{b}{Equivalent Width}

\tablenotetext{c}{Total $\phpcmsqps$ in the line from the HETGS observations by Ogle \etal~(2000)}

\end{deluxetable}

\vfil\eject\clearpage
\begin{deluxetable}{ccccccccc}
\tabletypesize{\footnotesize}
\tablewidth{0pt}
\rotate 
\tablecaption{Continuum Spectral Fits to the Extended Regions
  From the 3.2 s Frame Time Observation  \label{tab:ext_bremss}}
\tablecolumns{9} \tablehead{\colhead{Model\tablenotemark{a}} & \colhead{Region} &
  \colhead{$N_{H}$ (Soft Comp.)} & \colhead{kT\tablenotemark{b} / $\Gamma_{1}$ } & 
  \colhead{$K_{1}$\tablenotemark{c}} & \colhead{$N_{H}$ (Hard Comp.)} & 
  \colhead{ $\Gamma_{2}$}&
  \colhead{$K_{2}$\tablenotemark{c}} &
  \colhead{\chisq~/ d.o.f.}  \\ 
  \colhead{} & \colhead{} & \colhead{[$\times 10^{20}\pcmsq$]} & 
  \colhead{ } & \colhead{} & \colhead{[$\times 10^{22}\pcmsq$]} & \colhead{} 
   & \colhead{} & \colhead{} } \startdata

I & NE & 2.19 (frozen)  & $0.68^{+0.21}_{-0.12}$ &
$\left( 5.0^{+1.0}_{-1.3} \right) \times 10^{-5}$ & 2.6 (frozen) & 0.32 (frozen)  &
$\left( 6.2^{+1.5}_{-1.4} \right) \times 10^{-6}$ & 27.3/33\\

II & NE &  2.19 (frozen)  & $2.5^{+0.2}_{-0.3}$ & $\left(1.4^{+0.3}_{-0.2} \right) \times 10^{-5}$ &
3.1 (frozen)  & 0.32 (frozen) & $\left( 5.5^{+0.5}_{-0.4} \right) \times 10^{-6}$ & 32.0/33\\  

\hline

I & SW & 2.19 (frozen) & $0.41^{+0.04}_{-0.04}$ &
$\left( 1.8^{+0.4}_{-0.2} \right) \times 10^{-4}$ & 2.6 (frozen) & 0.32 (frozen)&
$\left( 6.2^{+1.4}_{-1.3} \right) \times 10^{-6}$ & 43.4/48\\

II & SW & 2.19 (frozen) & $3.2^{+0.1}_{-0.2}$ & $ \left( 2.0^{+0.4}_{-0.2} \right ) \times 10^{-5} $ & 
3.1 (frozen) & 0.32 (frozen) & $\left( 5.6^{+1.5}_{-1.3} \right) \times 10^{-6}$ & 54.2/48 \\

\enddata

\tablenotetext{a}{Model: I. bremsstrahlung + power law + emission lines;
			II. power law + power law + emission lines. (See Table 6 for emission line parameters.) }
\tablenotetext{b}{Unit: keV}
\tablenotetext{c}{Model normalization. 
  For the bremsstrahlung model $K_{\rm Brem} = 3.02 \times 10^{-15}
 \int n_e n_I dV / (4 \pi D^2)$, where $n_e$ is the electron density, $n_I$ is the ion
  density and $D$ is the distance to the source (all in cgs units).  For the power law model, $K_{\rm PL}
  = \phpcmsqps {\rm keV}^{-1}$ at 1 keV. }

\end{deluxetable}

\vfil\eject\clearpage
\begin{deluxetable}{ccccccc}
\tabletypesize{\footnotesize}
\tablewidth{0pt}
\rotate
\tablecaption{Emission Lines From the Extended Regions}
\label{tab:ne_lines}
\tablecolumns{7} \tablehead{\colhead{Model\tablenotemark{a}} & \colhead{Region} & \colhead{Energy} &
  \colhead{Line} & \colhead{Observed energy} &
  \colhead{K\tablenotemark{b}} & \colhead {EW} \\ 
\colhead{} & \colhead{} & \colhead{[keV]}
  & \colhead{} & \colhead{[keV]} & \colhead{} & \colhead{[eV]} } \startdata

I & NE & 0.90, 0.91, 0.92 \& 1.02 & Ne {\sc ix}  triplet, Ne {\sc x} Ly$\alpha$ & $0.96^{+0.04}_{-0.05}$ & $\left(
  3.1^{+1.0}_{-1.8} \right) \times 10^{-6}$ & 160.\\

I & NE & 0.74 \& 0.77 &O {\sc vii} RRC, O {\sc viii} Ly${\beta}$ & $0.72^{+0.03}_{-0.04}$ & $\left(
  3.9^{+1.9}_{-1.7} \right) \times 10^{-6}$ & 101. \\

I & NE & 0.561, 0.568 \& 0.574 & O {\sc vii} triplet & $0.58^{+0.02}_{-0.02}$ & $\left(
  8.4^{+4.2}_{-4.2} \right) \times 10^{-6}$ &  127.\\
\hline

II & NE & 0.90, 0.91, 0.92 \& 1.02 & Ne {\sc ix}  triplet, Ne {\sc x} Ly$\alpha$ & $0.95^{+0.04}_{-0.03}$ & $\left(
  4.0^{+0.9}_{-1.0} \right) \times 10^{-6}$ & 252\\

II & NE & 0.74 \& 0.77 &O {\sc vii} RRC, O {\sc viii} Ly${\beta}$ & $0.72^{+0.03}_{-0.02}$ & $\left(
  5.3^{+2.3}_{-2.2} \right) \times 10^{-6}$ & 172. \\

II & NE & 0.561, 0.568 \& 0.574 & O {\sc vii} triplet & $0.58^{+0.02}_{-0.02}$ & $\left(
  1.0^{+0.5}_{-0.4} \right) \times 10^{-5}$ &  194.\\

\hline
\hline

I & SW & 0.90, 0.91, 0.92 \& 1.02 & Ne {\sc ix}  triplet, Ne {\sc x} Ly$\alpha$ & $0.93^{+0.02}_{-0.04}$ & $\left(
  4.2^{+0.6}_{-0.9} \right) \times 10^{-6}$ & 124.\\

I & SW & 0.74 \& 0.77 &O {\sc vii} RRC, O {\sc viii} Ly${\beta}$ & $0.74^{+0.04}_{-0.06}$ & $\left(
  6.0^{+1.7}_{-2.1} \right) \times 10^{-6}$ & 86.6 \\

I & SW & 0.561, 0.568 \& 0.574 & O {\sc vii} triplet & $0.58^{+0.02}_{-0.01}$ & $\left(
  2.6^{+0.6}_{-0.9} \right) \times 10^{-5}$ &  178.\\

\hline

II & SW & 0.90, 0.91, 0.92 \& 1.02 & Ne {\sc ix}  triplet, Ne {\sc x} Ly$\alpha$ & $0.92^{+0.02}_{-0.03}$ & $\left(
  5.5^{+1.5}_{-1.6} \right) \times 10^{-6}$ & 221.\\

II & SW & 0.74 \& 0.77 &O {\sc vii} RRC, O {\sc viii} Ly${\beta}$ & $0.72^{+0.04}_{-0.03}$ & $\left(
  8.8^{+2.8}_{-2.3} \right) \times 10^{-6}$ & 180. \\

II & SW & 0.561, 0.568 \& 0.574 & O {\sc vii} triplet & $0.58^{+0.01}_{-0.01}$ & $\left(
  3.2^{+0.5}_{-0.6} \right) \times 10^{-5}$ &  284.\\

\enddata

\tablenotetext{a}{Models: See Table 5.} 
\tablenotetext{b}{$K = {\rm total} \phpcmsqps$ in the
  line.}

\end{deluxetable}

\vfil\eject\clearpage

\figcaption[fig1.ps] {Grey scale representations of {\it Chandra} X-ray
 images of NGC~4151 in the 0.3--2.5 keV  band taken with various frame time. 
 The narrow streak running NE--SW (P.A. $27 \degmark$) across the entire field, and visible
 in (a) and (b), is not real but a result of the CCD readout
 of the intense nuclear source. 	  
 (a) 0.1~s frame time. The grey scale is proportional to the square root of the X-ray
 intensity, ranging from 0 cts pixel$^{-1}$ (white) to 100 cts pixel$^{-1}$ (black).    
 (b) 0.4~s frame time.  The grey scale is proportional to the square root of the X-ray
 intensity, ranging from 0 cts pixel$^{-1}$ (white) to 100 cts pixel$^{-1}$ (black).
 (c) 3.2~s frame time, the grey scale is proportional to the square root of the X-ray
 intensity, ranging from 0 cts pixel$^{-1}$ (white) to 71 cts pixel$^{-1}$ (black).
 The ``hole'' at the nucleus 
 in the 3.2~s frame time image is not real but caused by pile up. 
\label{fig:im_grey}}

\figcaption[fig2.ps] {Contour plot of the 3.2~s frame time {\it Chandra} X-ray image of 
 NGC~4151 in the 0.3--10 keV range. The 11 contour levels range from 4 to 179 cts pixel$^{-1}$ and 
 are proportional to the square root of the X-ray intensity. 
  The hole in the nucleus is not real but a result of pile up.   
\label{fig:xcontour}}

\figcaption[fig3.ps] {Comparison of model point spread functions (PSFs) with observed radial profiles. 
     In each panel, the histogram is the model PSF from the PSF library and is normalized 
     to the 0.1~s frame time image peak, which is not piled-up. The points 
     are the observed count rates per pixel which were obtained 
     by averaging over an annulus or segments of an annular region of radial extent $1 \arcsec$. 
     Diamonds represent 0.1~s frame-time data, triangles represent  0.4~s frame-time data and squares represent 
     3.2~s frame-time data.  (a) The 2 -- 9 keV band. The PSF was interpolated  
     to 5.5 keV from the library of PSFs. The observed counts were taken from complete, circular annuli and are 
     consistent with the PSF, so 
     the emission in this band is unresolved (b) The 0.4 -- 2 keV band. 
     The PSF was interpolated  to 1.2 keV. The observed counts were taken from  
     two pie shaped regions each with opening angle $60 \degmark$ along P.A. $144\degmark$ and P.A. $324\degmark$, 
     which are approximately perpendicular to the direction of elongation of 
     the extended emission (see Fig. 1). This diagram shows that the soft emission along these 
     directions is marginally resolved 
     or unresolved. (c) Same energy range as (b). The counts were taken from a pie shaped 
     region toward the SW (P.A. $233\degmark$) with opening angle of 
     $120 \degmark$. (d) Same as (c) but for the NE region (P.A. $54\degmark$), with opening angle $120\degmark$.
\label{fig:psfvsobs}}

\figcaption[fig4.ps] {X-ray spectra of the nucleus of NGC~4151, 
  extracted from the 0.1~s and 0.4~s frame-time data. The upper panels show the
  data points with error bars (crosses), with the model folded through
  the instrument response (solid line). The lower panels show the $\chi$ residuals from this fit.
  The parameters of these models are listed in Tables 2, 3 and 4. Note that the calibration is uncertain
  below 0.50 keV and degrades rapidly below 0.45 keV.
  (a) The 0.1~s frame time data with a model comprising two power law continua plus emission lines (Tables 2 and 3).
  (b) The 0.1~s frame time data with a model comprising a bremsstrahlung and a power-law continuum, plus emission lines (Tables 2 and 3).
  (c) The 0.4~s frame time data with a model comprising a bremsstrahlung and a power-law continuum, plus emission lines (Table 4). The 
      channels above 7 keV are ignored in the fit to reduce the effect of ``pile-up''.
\label{fig:spec_nuc}}

\figcaption[fig5.ps] {The rectangles defining the SW and NE regions are shown superposed on a grey 
scale representation of the 3.2 s frame-time 0.3 -- 10 keV image.  Regions within the white circle suffer from pile-up. 
\label{fig:extract}}

\figcaption[fig6.ps] {Hardness ratio (defined as the ratio of the count rates
 in the 0.8--2 keV to the 0.4--0.8 keV band) as a function of distance from 
 the nucleus along P.A. $233\degmark$. The hardness ratio was obtained from the 
 3.2~s frame time data. These data suffer from pile-up at $r \leq 3\arcsec$. 
\label{fig:hardness}}

\figcaption[fig7.ps] {X-ray spectra of the extended regions of NGC~4151
  extracted from the 3.2~s frame-time data. The upper panels show the
  data points with error bars (crosses) and the model folded through
  the instrument response (solid line). The lower panels show the $\chi$ residuals to this fit.
  The parameters of the models are listed in Tables 5 and 6.
  Note that the calibration is uncertain below
  0.50 keV and degrades rapidly below 0.45 keV.
  (a) The SW region. The model is a bremsstrahlung plus a power law plus emission lines.
  (b) The SW region. The model is two power laws plus emission lines.	
  (c) The NE region. The model is a bremsstrahlung plus a power law plus emission lines.
  (d) The NE region. The model is two power laws plus emission lines.
\label{fig:spec_ne}}

\figcaption[fig8.ps] {A superposition of an [OIII] $\lambda$5007 image (contours, from
 P\'{e}rez-Fournon \& Wilson 1990) on
 the 3.2~s frame time X-ray image. The grey scale is proportional to the square root of the X-ray
 intensity, ranging from 1 cts pixel$^{-1}$ (white) to 40 cts pixel$^{-1}$ (black). 
 The contour levels for the optical image are drawn at 0.07\%, 0.15\%, 0.3\%, 0.6\%, 0.12\%, 2.5\%, 5\%, 
 10\%, 20\% and 80\% of the peak. 
\label{fig:xray_OIII}}

\figcaption[fig9.ps] {A superposition of a MERLIN radio continuum image at 21 cm (contours, from Mundell \etal~1995)
 on the 0.1~s frame time grey scale X-ray image.  The grey scale is proportional to the square root of the 
 X-ray intensity, ranging from 0 cts pixel$^{-1}$ (white) to a peak intensity of 314 cts pixel$^{-1}$ (black).  
 The contour levels of the radio image are drawn at 1.5\%, 3\%, 5\%, 12\%, 20\%, 75\% of the peak.
\label{fig:xray_radio}}

\figcaption[fig10.ps] {X-ray spectrum of the off nuclear source described in
the Appendix. The spectrum is extracted from
the 3.2~s frame time observation. The upper panel shows the
data points with error bars (crosses) and the model folded through
the instrument response (solid line). The lower panel shows
the $\chi$ residuals to this fit. The model is an absorbed power-law. 
\label{fig:hotspot}}

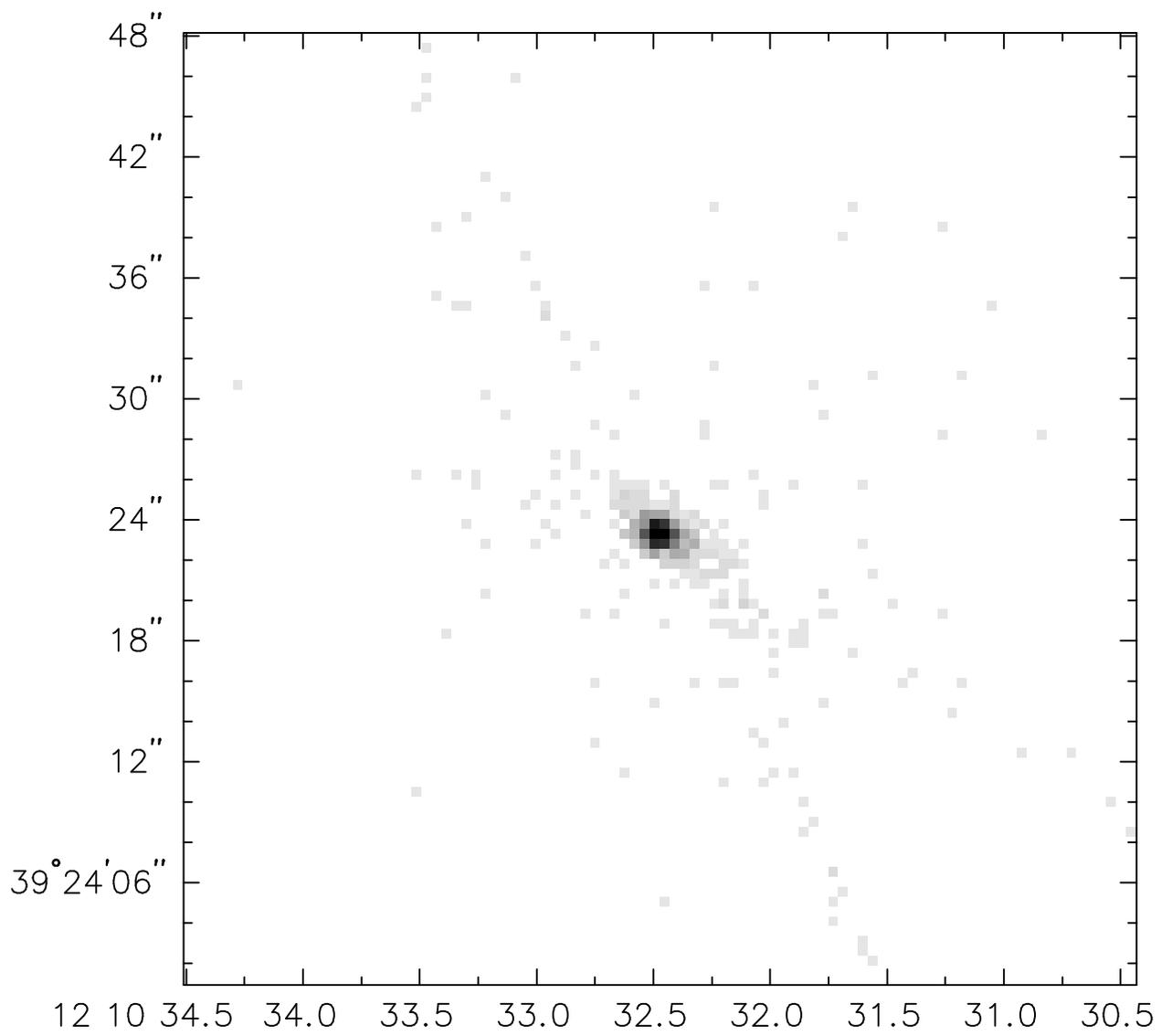
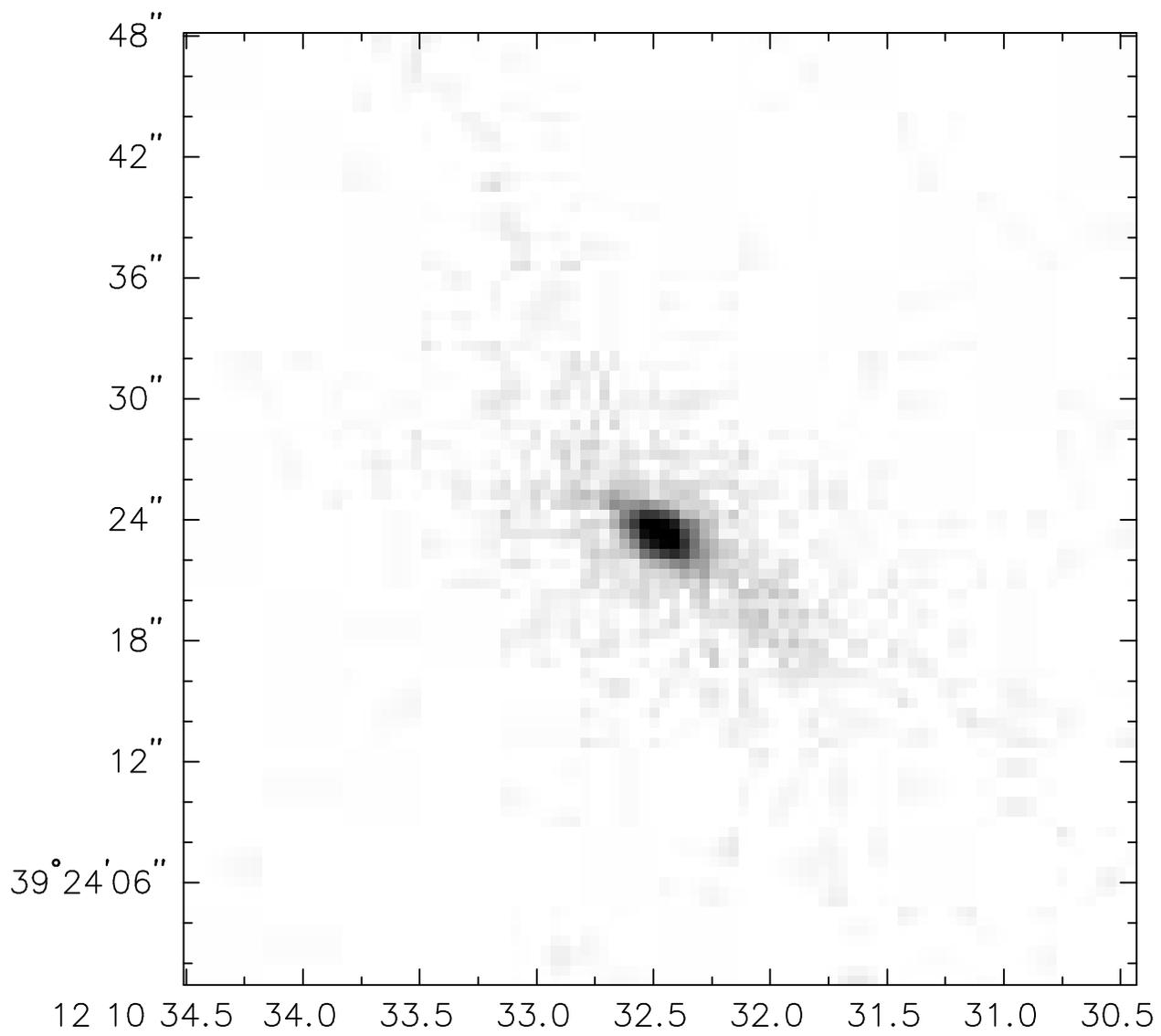
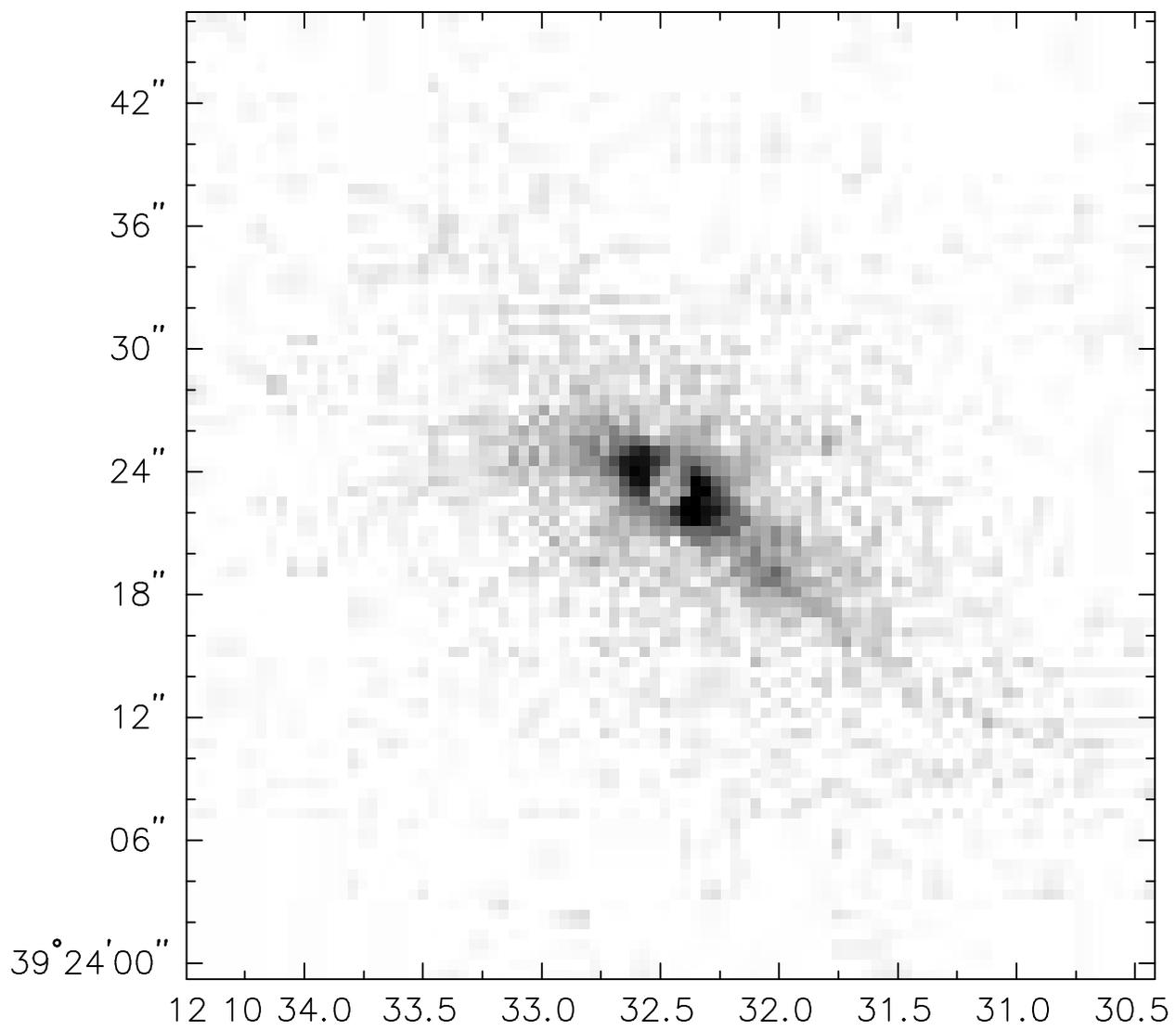
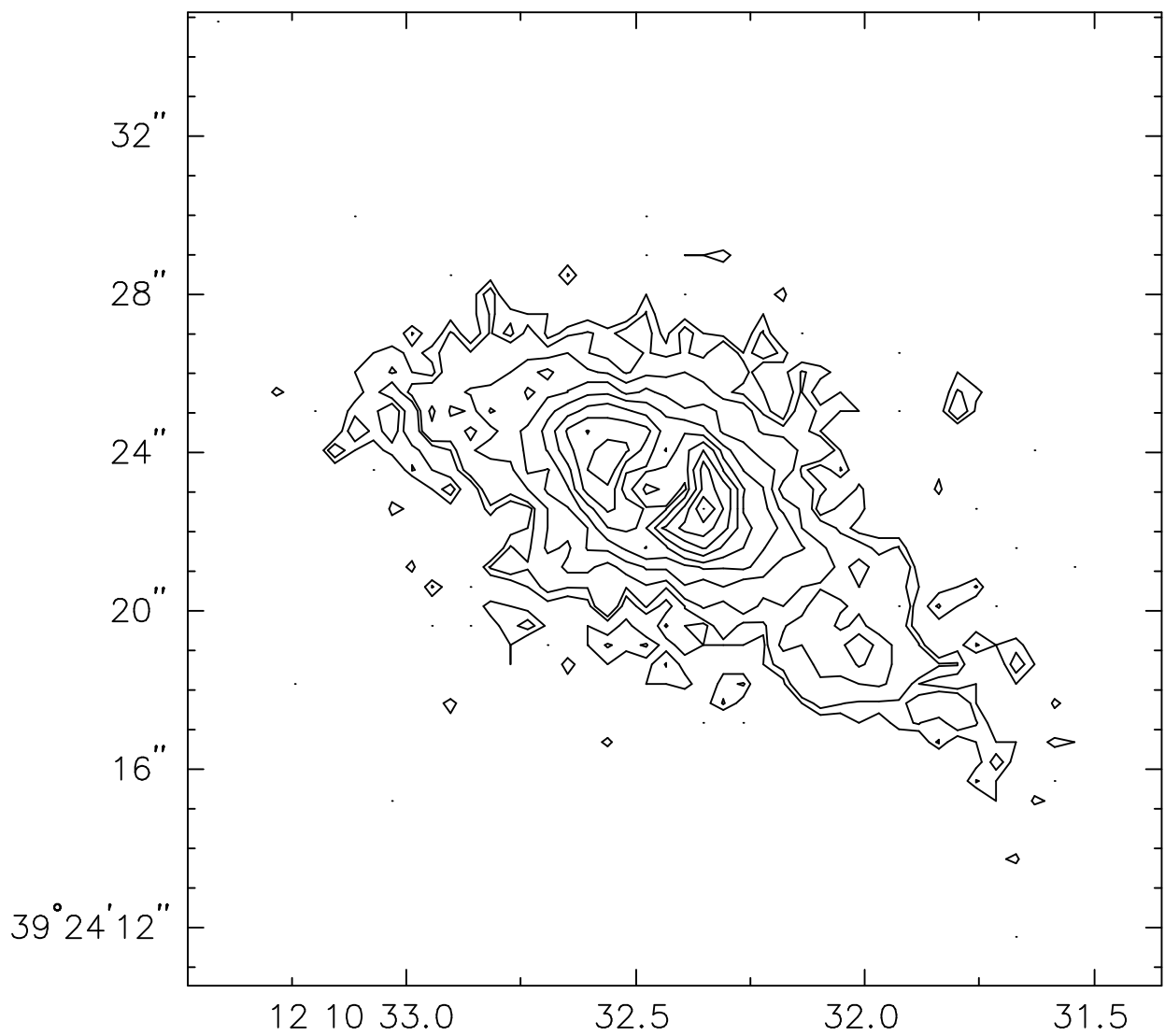
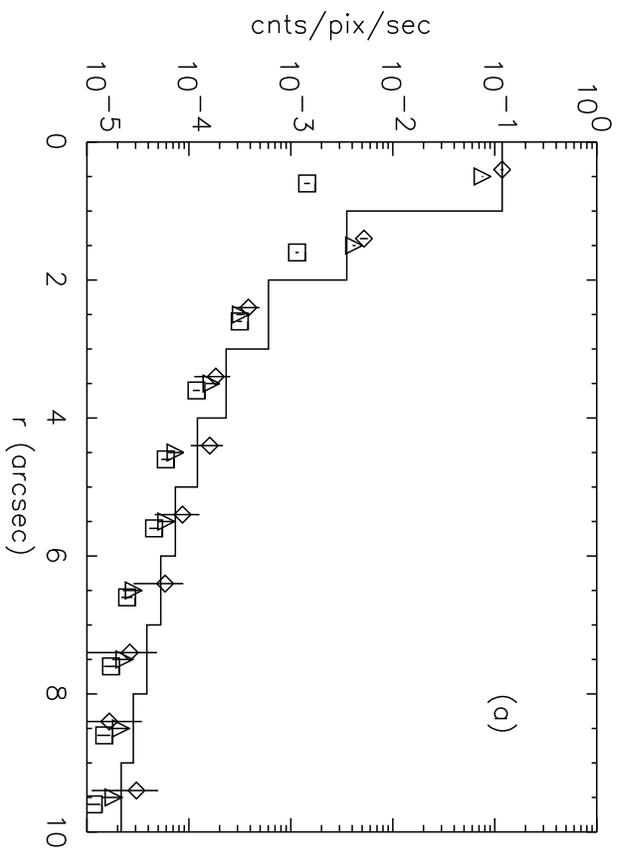
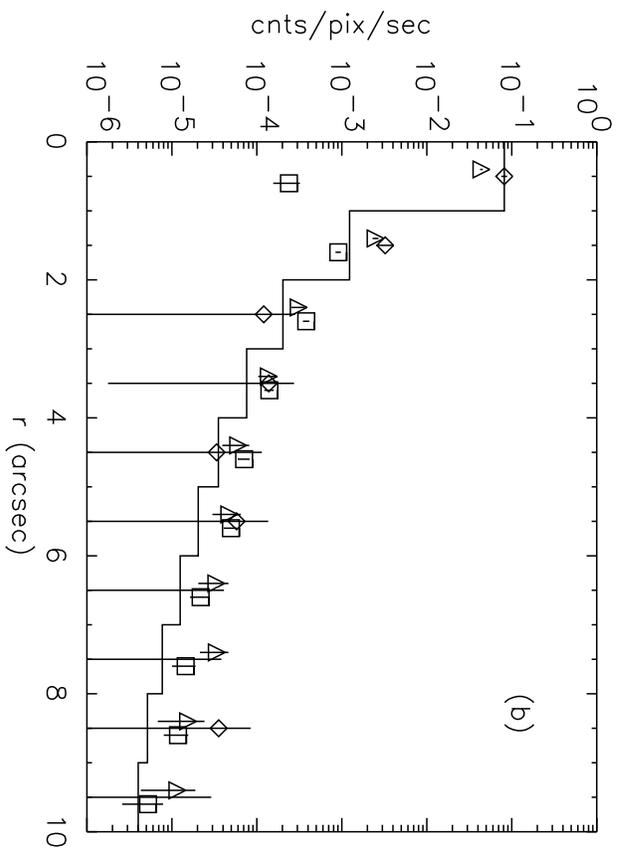
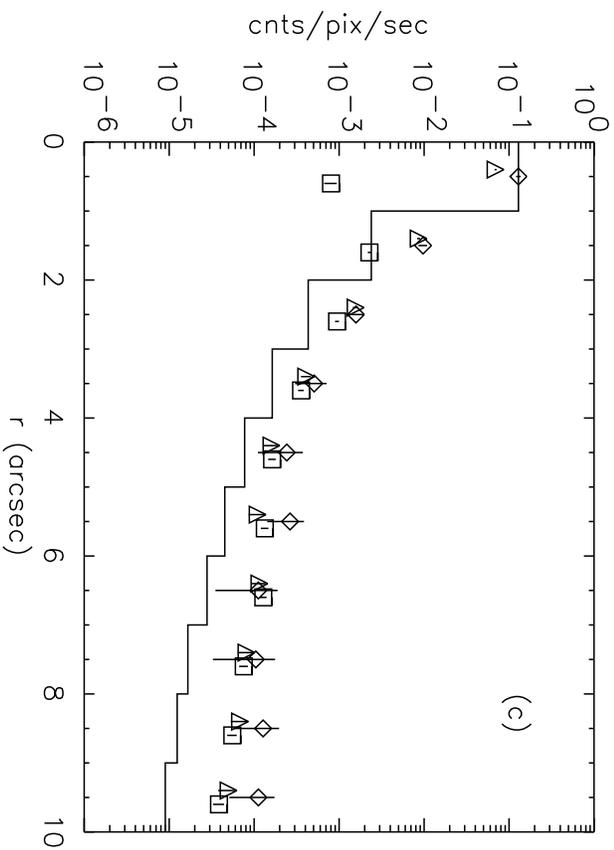
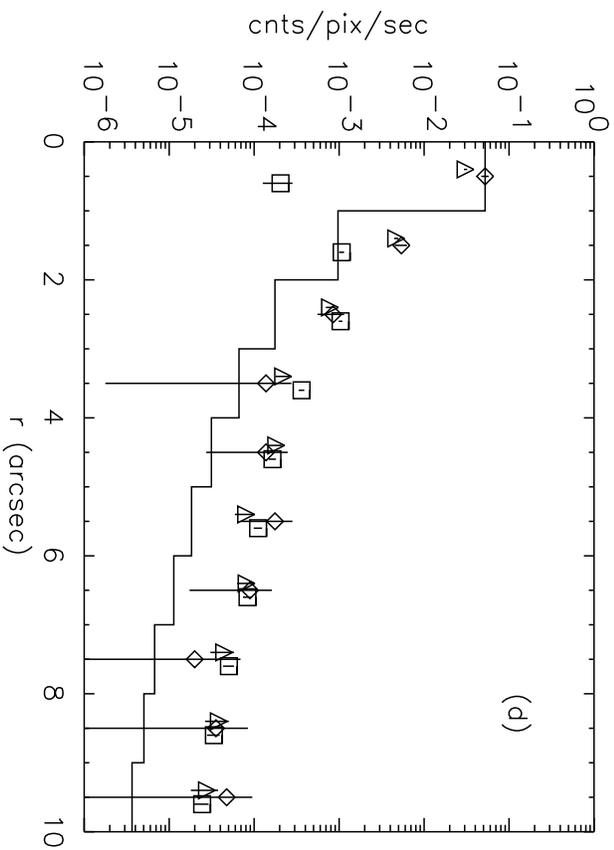
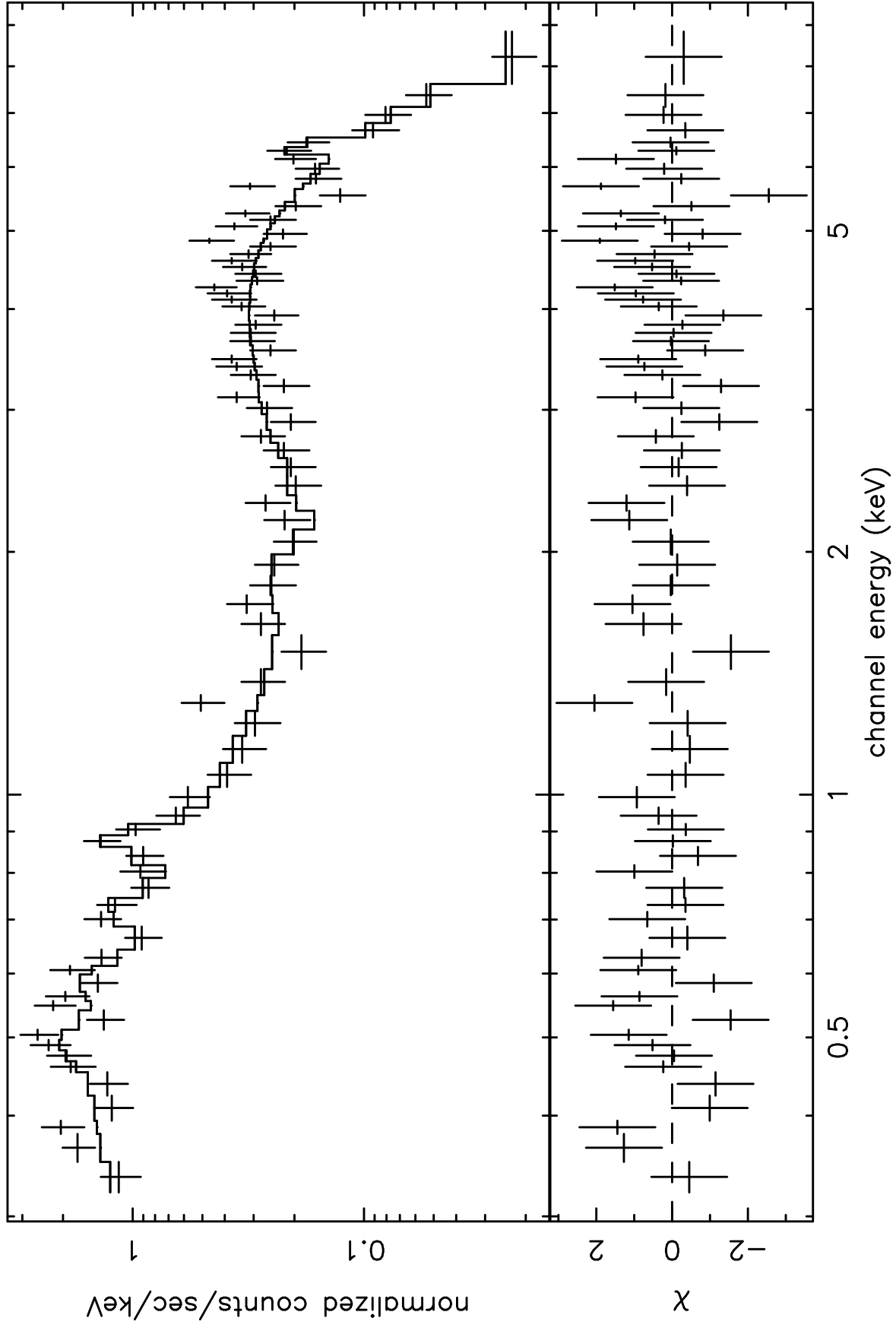
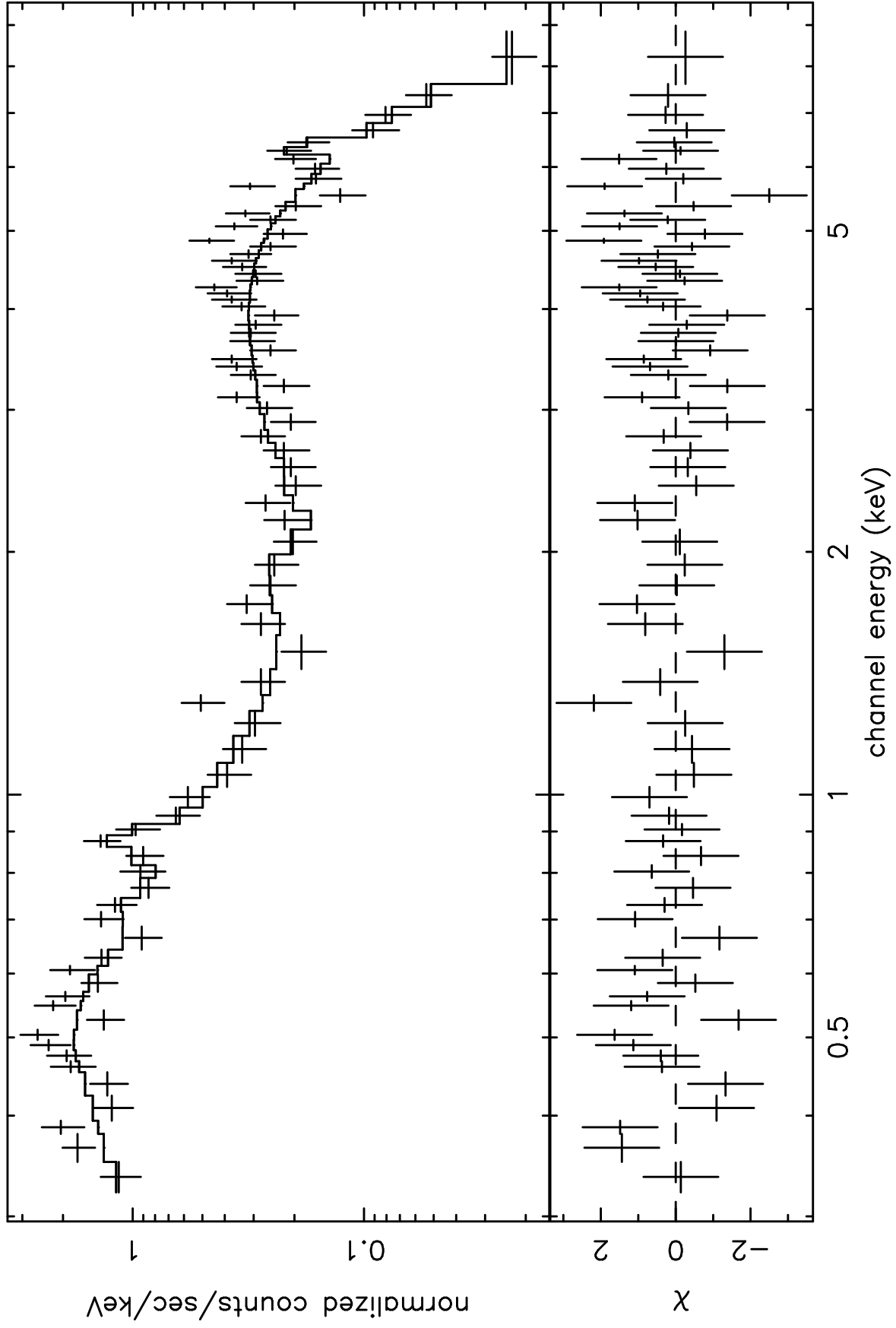
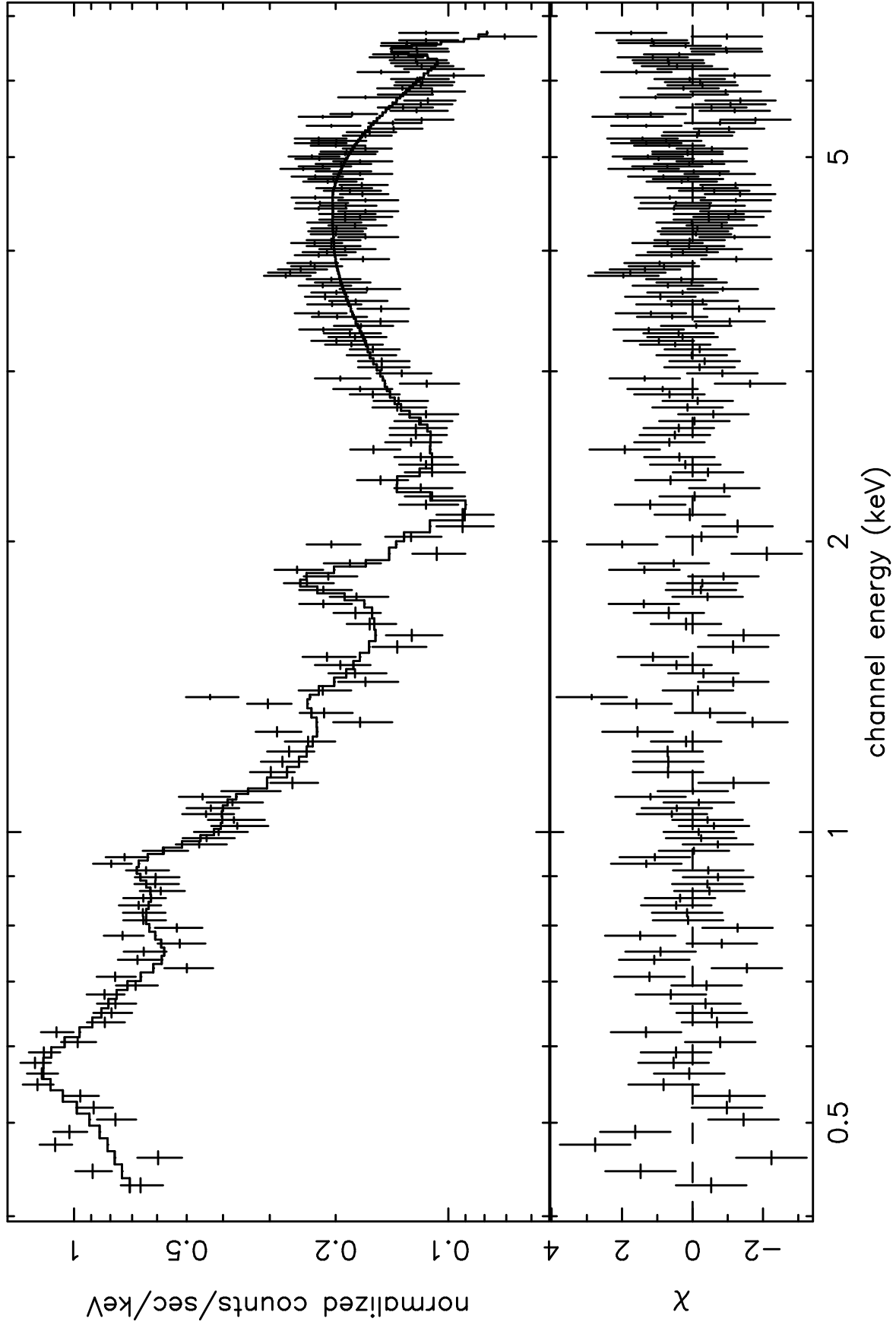
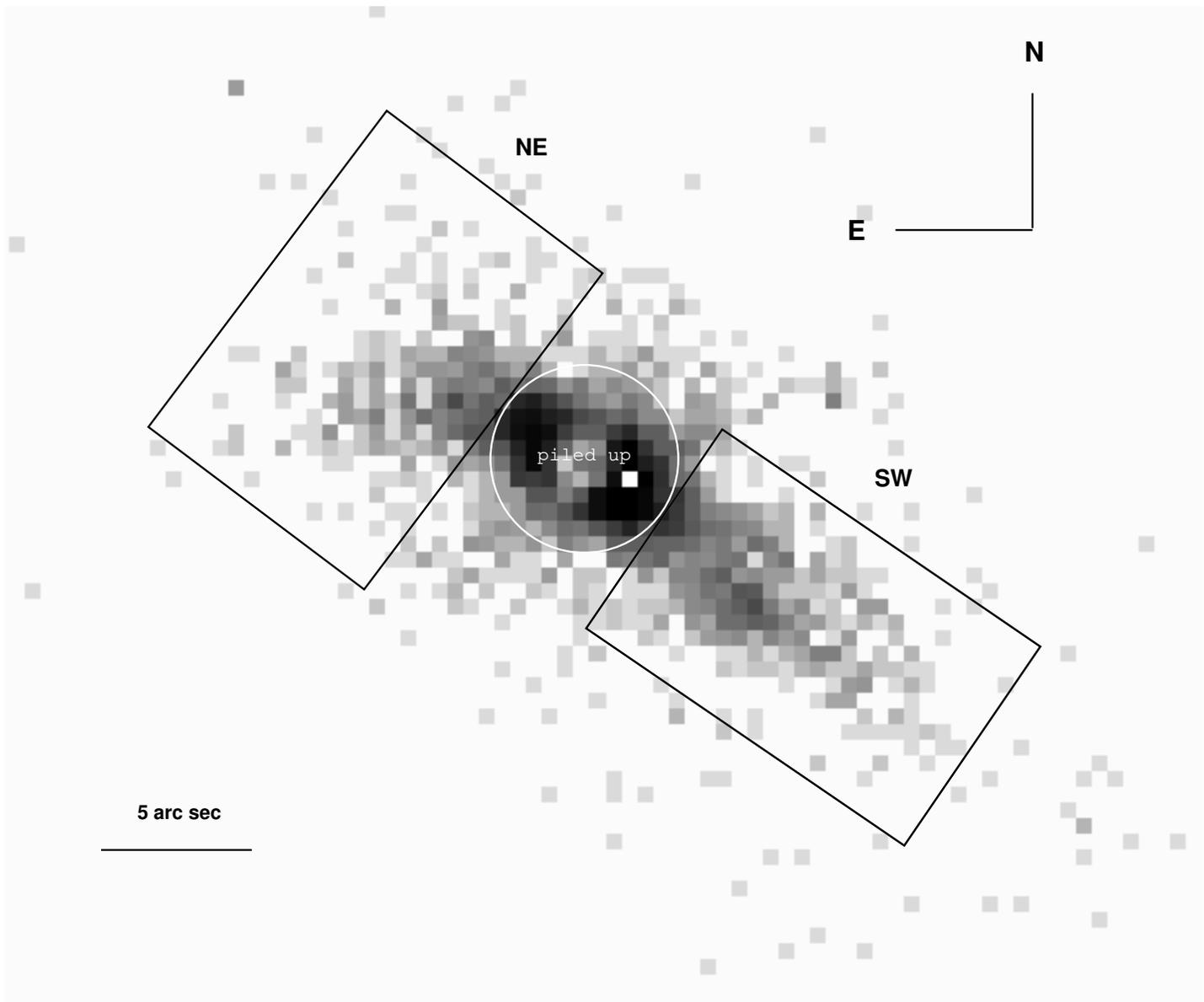
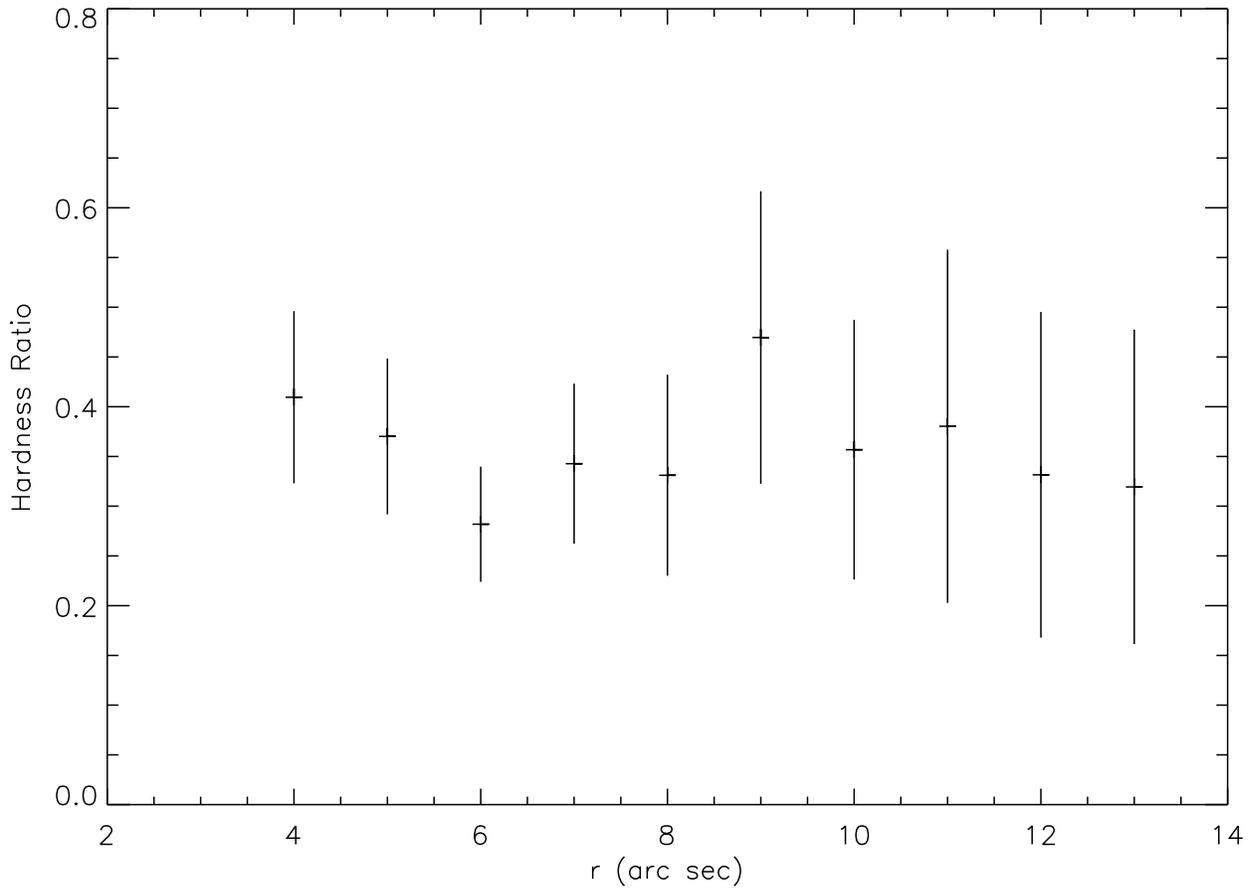
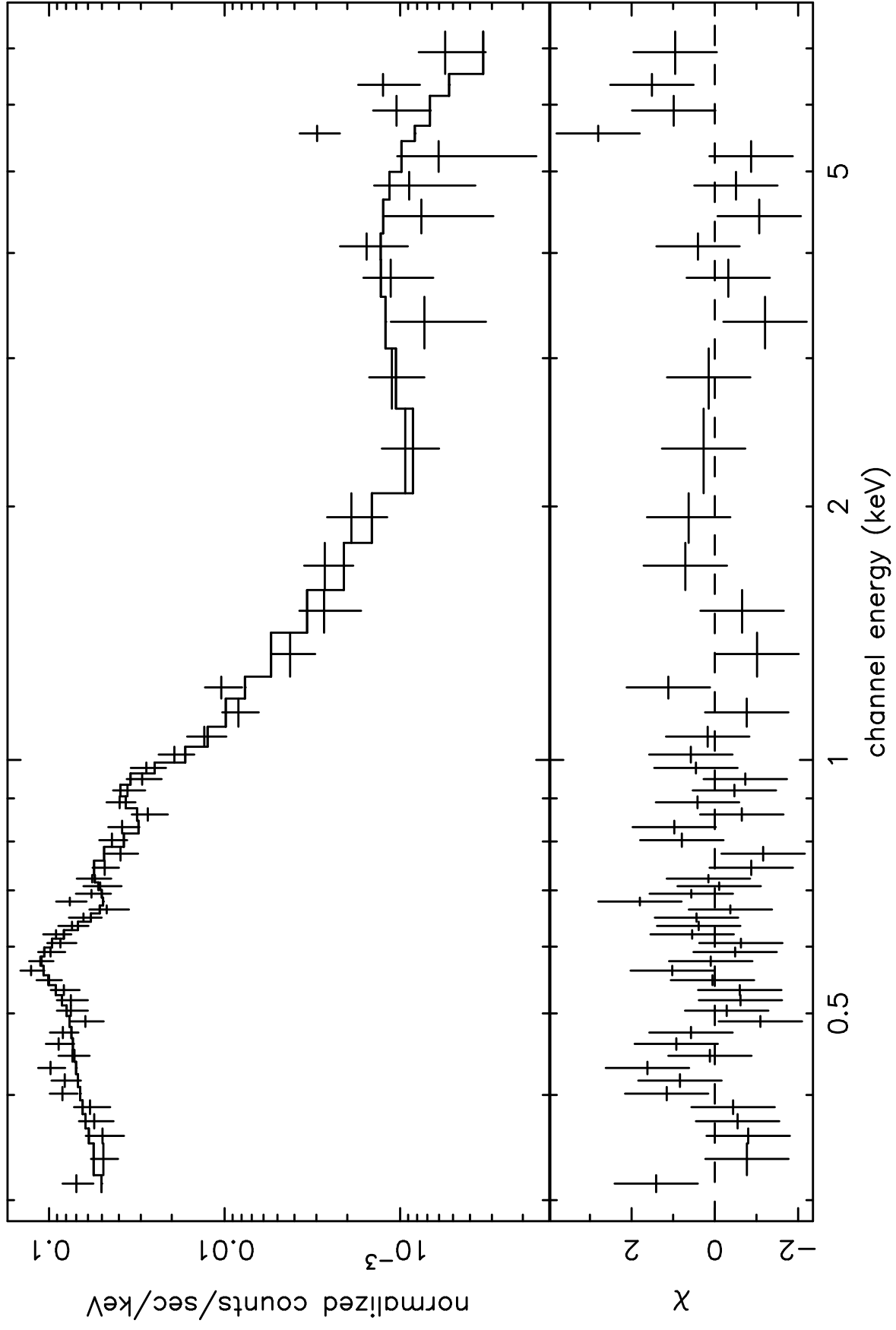
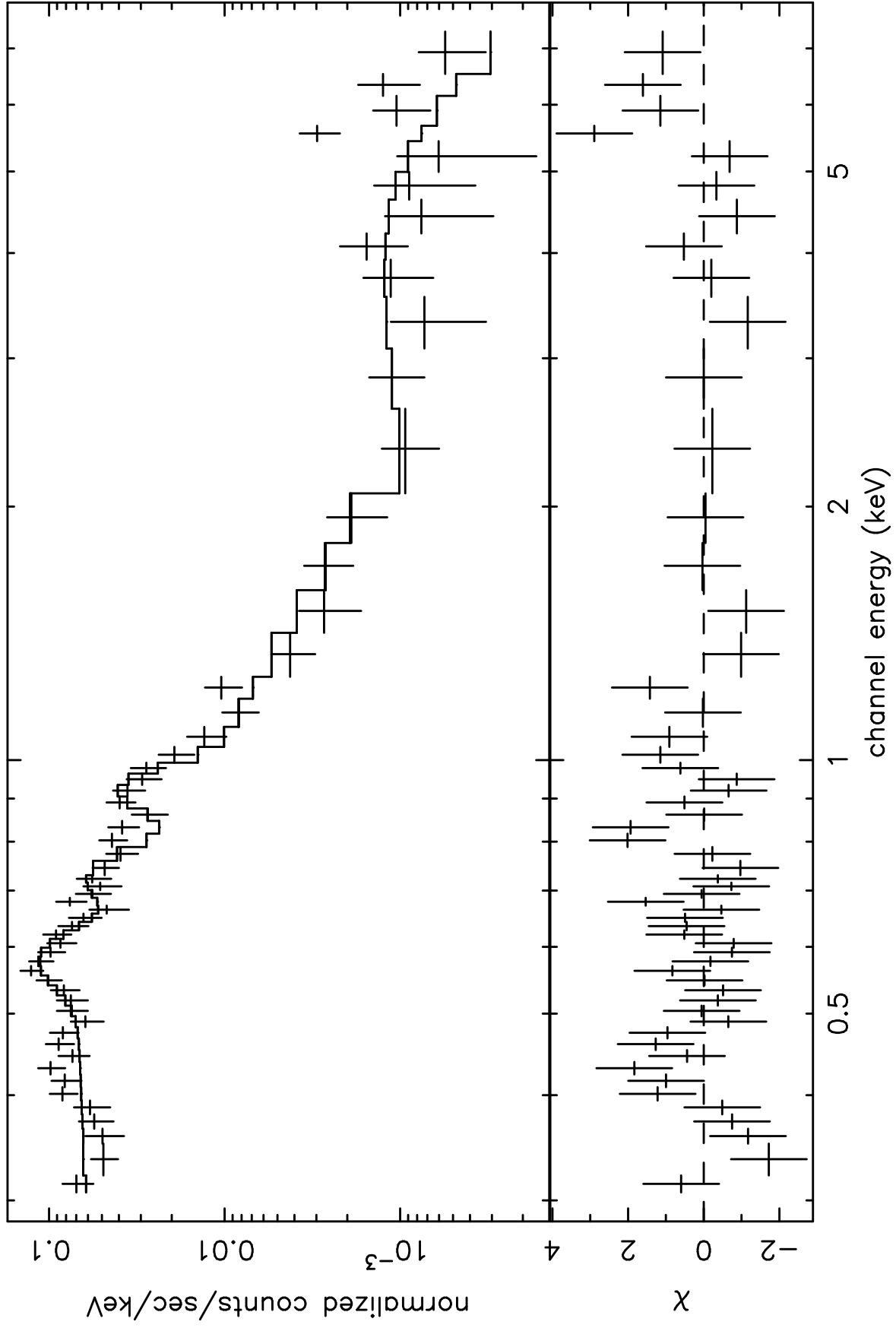
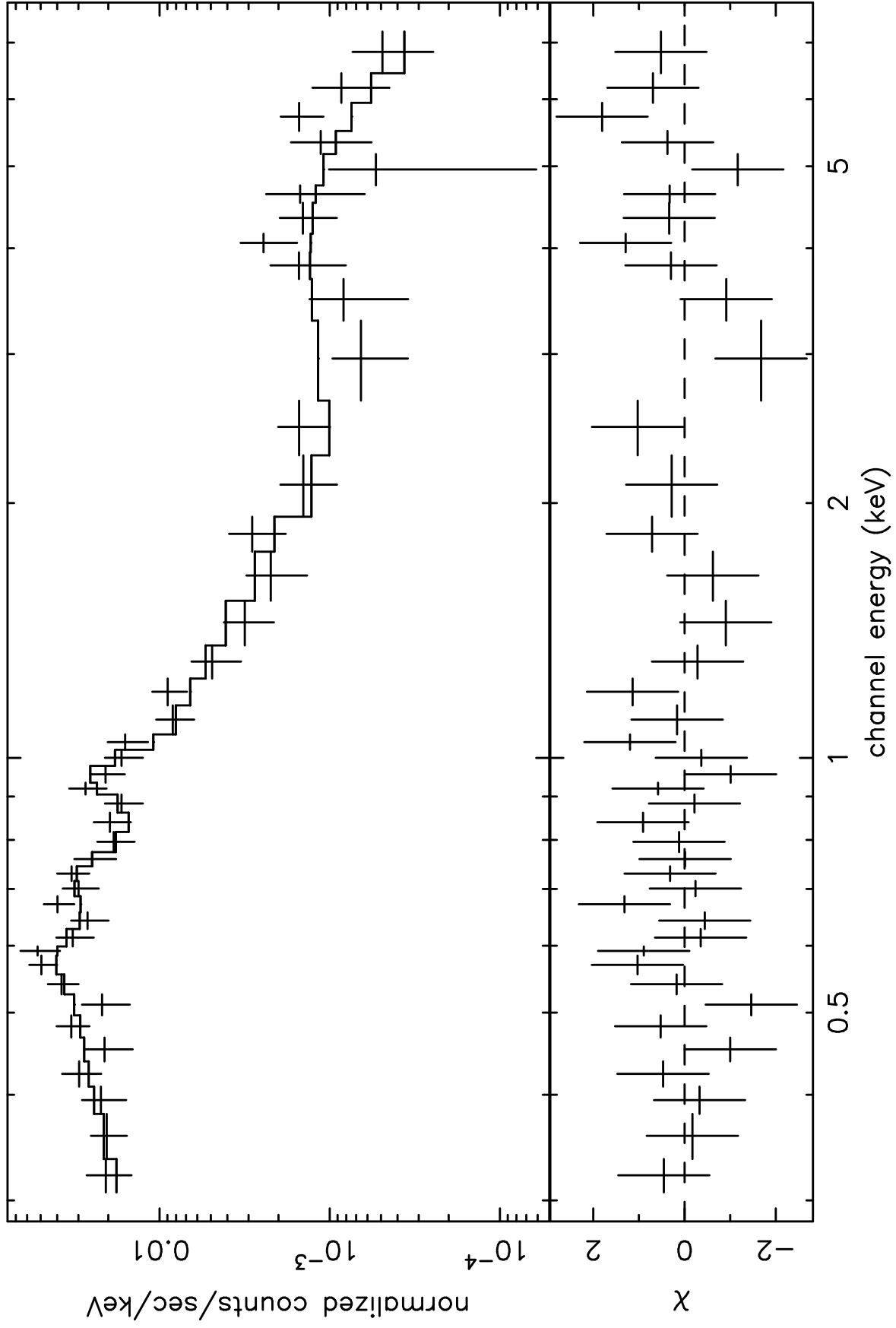
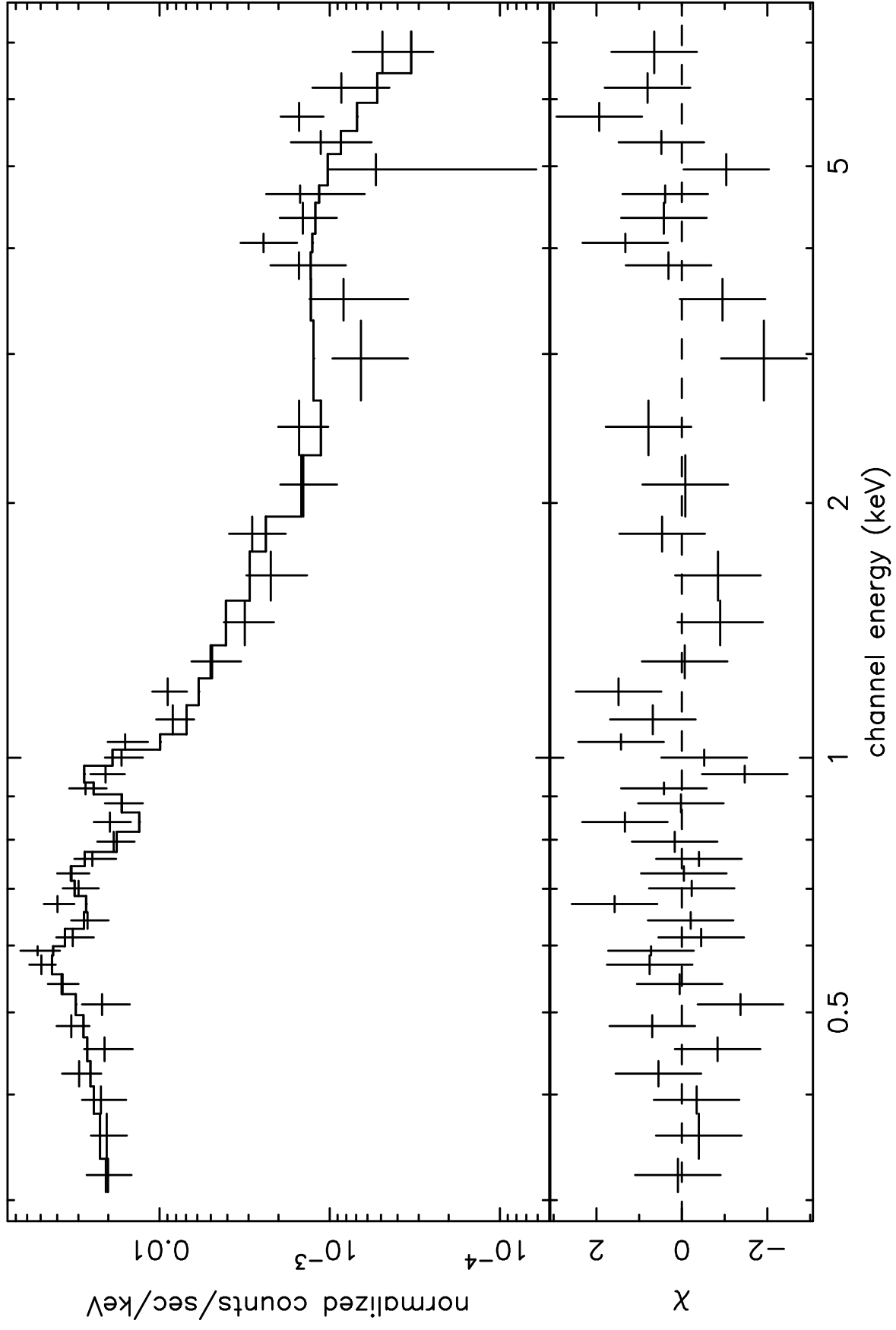
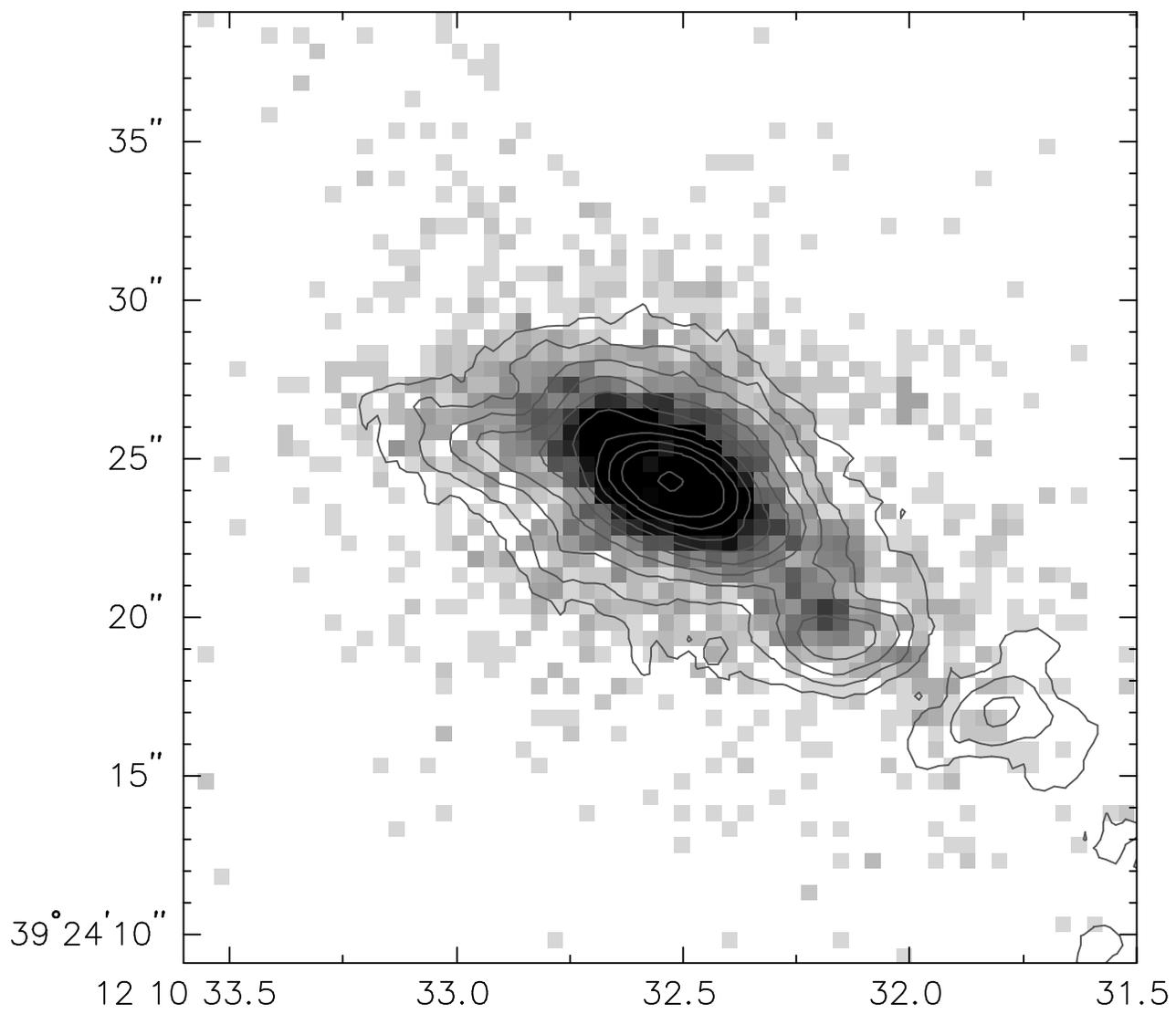
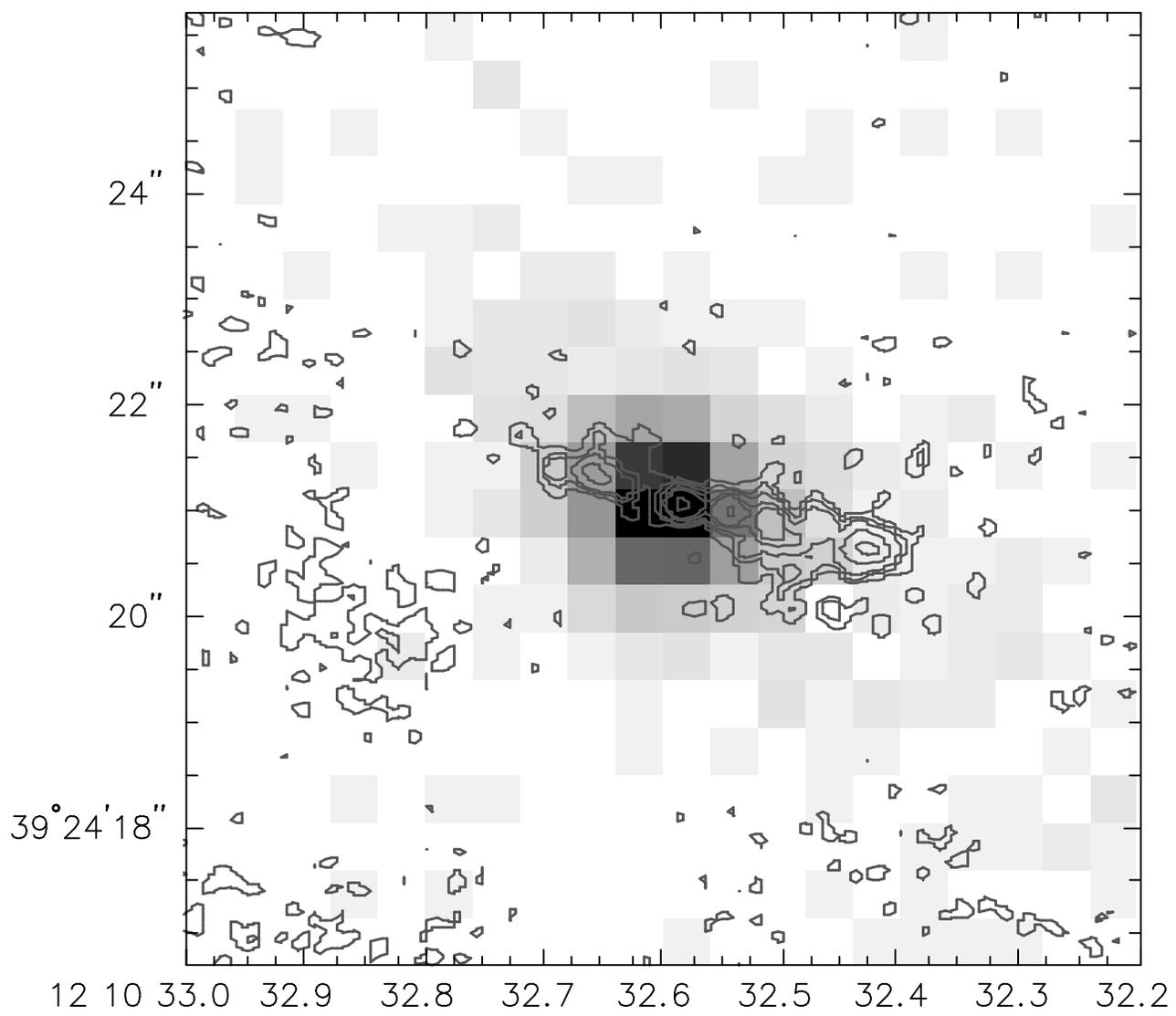
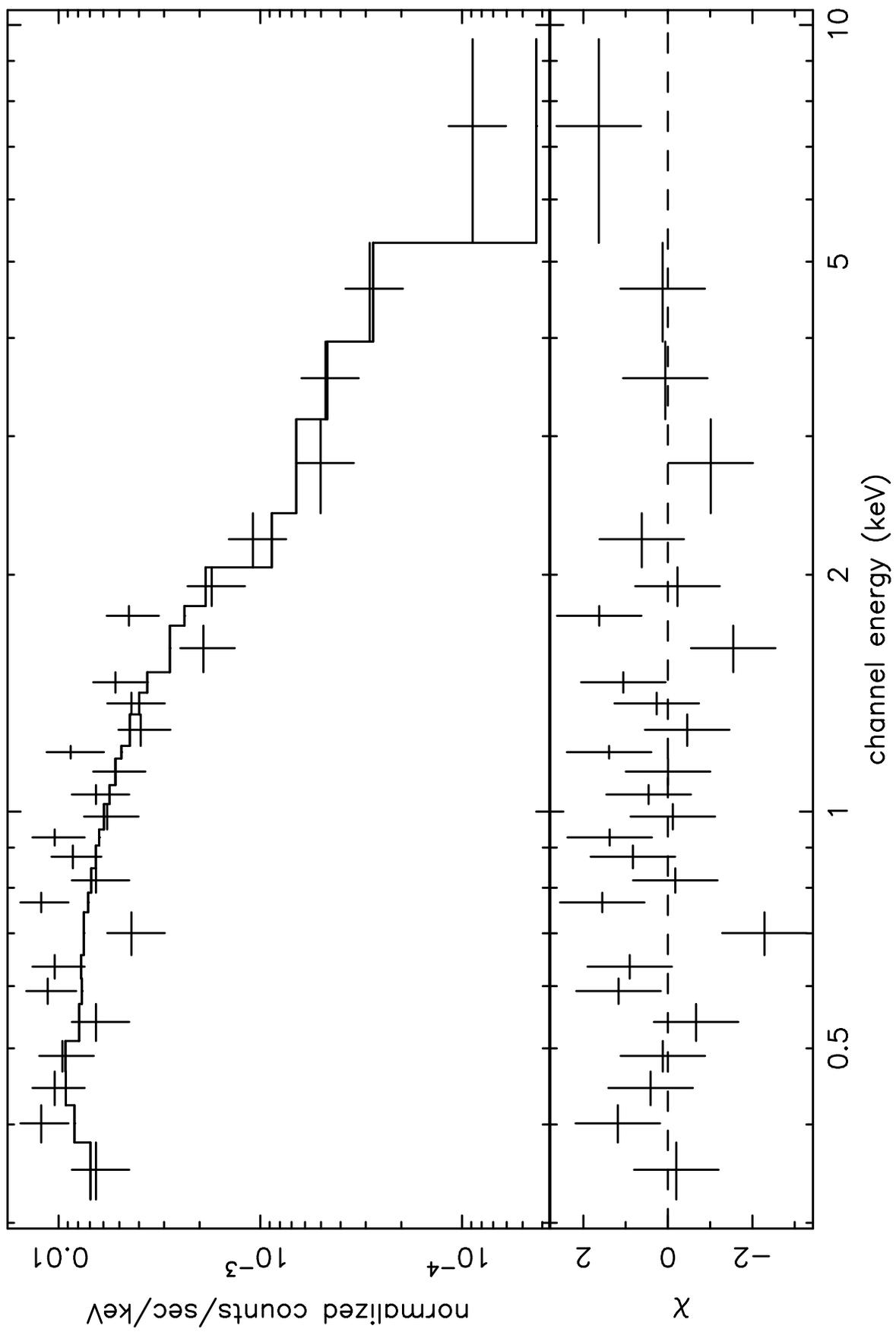
\end{document}